\advance\voffset by -20mm
\advance\hoffset by -15mm
\documentstyle[12pt]{article}
\begin{document}
\null
\vspace{30mm}

\noindent{\Large{\bf SOLITON SOLUTIONS OF\\[2mm]
RELATIVISTIC HARTREE'S EQUATIONS
}}\vspace{17mm}

\noindent{\large Nathan Poliatzky$^*$
}\vspace{5mm}

\noindent Institut f\"ur Theoretische Physik, ETH-H\"onggerberg,
CH-8093 Z\"urich, Switzerland
\vspace{28mm}

\noindent {\small {\em Abstract.}
We study a model based on $N$ scalar complex fields coupled to a
scalar real field, where all fields are treated classically as
c-numbers. The model describes a composite particle made up of $N$
constituents with bare mass $m_0$ interacting both with each other and
with themselves via the exchange of a particle of mass $\mu_0$. The
stationary states of the composite particle are described by
relativistic Hartree's equations. Since the self-interaction is
included, the case of an elementary particle is a nontrivial special
case of this model. Using an integral transform method we derive the
exact ground state solution and prove its local stability. The mass of
the composite particle is calculated as the total energy in the rest
frame. For the case of a massless exchange particle the mass formula
is given in closed form. The mass, as a function of the coupling
constant, possesses a well pronounced minimum for each value of
$\mu_0/m_0$, while the absolute minimum occurs at $\mu_0=0$.
}\vspace{17mm}

\section{Introduction and results}\label{Introduction}
In this paper we derive and study the exact ground state solution in
$3+1$ dimensions for a self-interacting system whose dynamics is
governed by the relativistic Hartree's equations. Our motivation is to
study bound states formed solely by self-interaction. Such bound states
(sometimes called nontopological solitons) offer a possibility of
understanding the internal properties of particles, such as their
masses, charges and magnetic moments. Self-interaction is a purely
nonlinear and nonperturbative phenomenon which is neither
well understood nor properly appreciated at present. However, we
argue that in the quantum domain it is rather important and deserves
close study. It may be the crucial missing element in our understanding
of quantum phenomena. Self-interaction appears frequently in quantum
field theories such as, for example, in quantum electrodynamics, where
it leads to infinities. Intuitively, the reason why these infinities
appear is that we are using a perturbation theory in which particles
are associated with free fields such as plane
waves or wave packets. The fact that one is able to cure the theory and
remove the infinities in a self-consistent manner can be seen as an
indication that the original theory, before the perturbation theory based
on free fields is applied to it, is correct and should merely be treated
differently. What is needed, perhaps, is an approach which is
based on the self-interaction bound states in place of free fields. An
attempt of such an approach for the case of quantum electrodynamics will
be presented in a forthcoming paper. In the present paper we attempt to
present the self-interaction bound states in their own right, within the
context of a classical field theory. In doing so we will concentrate on
their internal properties, in which they differ so much from free fields.

The following model is motivated by its simplicity, self-consistency and
by the fact that it is connected to the large $N_c$ limit of QCD. It
is not intended to be realistic in first place. Yet some of its
elements, for instance relativistic invariance, dimension $3+1$ and the
nonlinear interaction, are clearly realistic.

We consider a system of $N$ complex scalar fields $\Psi_j\left({\bf r},t
\right)$, $j=1,\ldots,N$ and a real scalar field $\Phi\left({\bf r},t\right)$
with the lagrangian
\begin{equation}{\cal L}=-\sum_{j=1}^N\left(\partial^\nu\Psi_j^\ast\partial_\nu
\Psi_j+m_0^2\Psi_j^\ast\Psi_j-g\Phi^p\left(\Psi_j^\ast\Psi_j\right)^q\right)-
\frac{1}{2}\left(\partial^\nu\Phi\partial_\nu\Phi+\mu_0^2\Phi^2\right)\,,
\label{lagr}\end{equation}
where $p=q=1$, and the equations of motion
\begin{equation}\left(\Box-m_0^2+g\Phi\left({\bf
r},t\right)\right)\,\Psi_j\left({\bf
r},t\right)=0\,,\label{eomPsi}\end{equation}
\begin{equation}\left(\Box-\mu_0^2\right)\,\Phi\left({\bf
r},t\right)=-g\sum_{i=1}^N\Psi_i^\ast\left({\bf r},t\right)
\Psi_i\left({\bf r},t\right)\,,\label{eomU}\end{equation}
where $\Box\equiv\triangle-\partial^2/\partial t^2\equiv\partial^\nu
\partial_\nu$. All the fields are assumed to be classical, i.e.
commuting. Notice that the interaction term in~(\ref{lagr}) is not positive
definite and hence the stability question arises. In this paper only
the local stability is investigated in full detail. Here it suffices
to say that stable one-particle and many-particle bound states exist,
with energies below the threshold to the continuum states, provided
certain restrictions are placed on the coupling constant $g$ and on
$\mu_0/m_0$. The fields $\Psi_j$ are normalized according to
\begin{equation}\int{\rm d}^3{\bf r}\left|\frac{\partial\Psi^\ast_j}{\partial
t}\Psi_j-\Psi^\ast_j\frac{\partial\Psi_j}{\partial t}\right|=1\,,\quad j=1,
\ldots,N\,.\label{normPsi}\end{equation}
Notice that the left-hand side of~(\ref{normPsi}) is a constant of
motion. In the present context the $1$ on the right-hand side of
(\ref{normPsi}) is purely conventional, in fact we could equally well
chose any other number. However, equations~(\ref{eomPsi}) and
(\ref{eomU}) allow for a rescaling of the fields, which can be chosen
to restore the $1$ in~(\ref{normPsi}). The fact, that $\Psi_j$ are
complex fields and hence are normalized according to~(\ref{normPsi}),
is an essential one since otherwise the model would contain the case
$\mu_0=m_0$, $\Psi=\Phi$ as a special case  and then would be
identical to the standard $\Phi^3$ model which is unstable even locally.

Equations~(\ref{eomPsi}-\ref{normPsi}) admit plane wave solutions with
continuous
energies $\vert E_{{\bf k}}\vert\equiv{\bf k}^2+m_0^2\geq m_0$
\begin{equation}\Psi_j\left({\bf r},t\right)=\frac{{\rm e}^{i{\bf k}_j\cdot
{\bf r}-iE_{{\bf k}_j}t}}{\sqrt{2\vert E_{{\bf k}_j}\vert\Omega}}\,,\quad \Phi
\left({\bf r},t\right)=\frac{g}{\mu_0^2}\frac{N}{2\vert E_{{\bf k}_j}\vert
\Omega}\,,\quad\Omega\rightarrow\infty\,,\label{plwavesol}\end{equation}
where $\Omega$ is an arbitrarily large normalization volume (a detailed
discussion of plane wave solutions in the context of nontopological
solitons is given in ref.~\cite{TDLee}). As will
be shown below, the characteristic feature of these states is that for
any $N$ their total energy is always positive and larger than the
corresponding bare mass $Nm_0$, i.e. $E_{\rm total}\ge Nm_0$ and hence
on the energy scale these states fill the continuum above $Nm_0$.

In contrast to the above continuum states there exist bound states with
total energies $0<E_{\rm total}<Nm_0$. In the rest frame of the bound
states the corresponding stationary state solutions of~(\ref{eomPsi}) and
(\ref{eomU}) are of the form $\Psi_j\left({\bf r},t\right)=\psi_j\left({\bf r}
\right)\,{\rm e}^{-iE_jt}$ and $\Phi\left({\bf r},t\right)=\,\phi\left({\bf r}
\right)/g$, where $\vert E_j\vert<m_0$.  For these stationary fields
the equations of motion~(\ref{eomPsi}),~(\ref{eomU}) and the
normalization condition~(\ref{normPsi}) become respectively
\begin{equation}\left(\triangle-\gamma_j^2+\phi\left({\bf r}\right)\right)\,
\psi_j\left({\bf r}\right)=0\,,\label{Hartreeeq}\end{equation}
\begin{equation}\left(\triangle-\mu_0^2\right)\phi\left({\bf r}\right)=-g^2
\sum_{i=1}^N\left|\psi_i\left({\bf r}\right)\right|^2\,,\label{Hartreepoteq}
\end{equation}
and
\begin{equation}\int{\rm d}^3{\bf r}\;\left|\,\psi_j\left({\bf r}\right)
\right|^2=\frac{1}{2\vert E_j\vert}\,,\quad j=1,\ldots,N\,,\label{normcond}
\end{equation}
where $\gamma^2_j=m_0^2-E_j^2\geq0$. Once a solution is obtained, it
can be Lorentz-boosted to an arbitrary inertial frame. Equations
(\ref{Hartreeeq}) and~(\ref{Hartreepoteq}) must be supplemented with
suitable boundary conditions (they are discussed below).

This system constitutes a model of a composite particle made up of $N$
spinless constituents of bare mass $m_0$ and the constituent energies
$\vert E_j\vert$, which interact with each other and with themselves via
the exchange of a particle of mass $\mu_0$. We identify the physical
mass $M$ of the composite particle with its total energy $E_{\rm
total}$. The total energy follows in the usual way from the lagrangian
(\ref{lagr})
\[M\equiv E_{\rm total}=\int{\rm d}^3{\bf r}\left(\sum_{j=1}^N\left[\nabla
\psi^\ast_j\cdot\nabla\psi_j+\left(E_j^2+m_0^2\right)\psi^\ast_j\psi_j-\phi
\psi^\ast_j\psi_j\right]+\frac{1}{2g^2}\left[\left(\nabla\phi\right)^2+\mu_0^2
\phi^2\right]\right).\]
The gradient term $\left(\nabla\phi\right)^2$ can be eliminated using
(\ref{Hartreepoteq}) and we obtain
\[M=\int{\rm d}^3{\bf r}\sum_{j=1}^N\left[\nabla\psi^\ast_j\cdot\nabla\psi_j+
\left(E_j^2+m_0^2\right)\psi^\ast_j\psi_j-\frac{1}{2}\phi\psi^\ast_j\psi_j
\right].\]
Eliminating the remaining gradient term by the aid of
(\ref{Hartreeeq}) and using the normalization condition
(\ref{normcond}), we get
\begin{equation}M=\sum_{j=1}^N\left(\vert E_j\vert+\frac{1}{2}\int{\rm d}^3
{\bf r}\;\phi\left({\bf r}\right)\,\left|\,\psi_j\left({\bf r}\right)\right|^2
\right)\,.\label{Etotal}\end{equation}
For the continuum states given in~(\ref{plwavesol}) the last term in
(\ref{Etotal}) vanishes as $\Omega\rightarrow\infty$ and hence $E_{\rm
total}=\sum_{j=1}^N\vert E_j\vert\equiv\sum_{j=1}^N\vert E_{{\bf
k}_j}\vert\geq Nm_0$. In the case of bound states the last term in
(\ref{Etotal}) does not vanish and hence the total mass is not
identical to the sum of the constituent energies $\vert E_j\vert$
(i.e., to the sum of the bare masses $m_0$ and the binding energies
$m_0-\vert E_j\vert$ of the constituents) but contains an additional
term. Below we will show that for stationary states this term is
strictly positive. While the binding energies can be interpreted as
the energies which are gained by putting particles in an attractive
potential, the last term in~(\ref{Etotal}) represents the energy which
must be spent to create the potential itself. In a consistent field
theory these two sorts of energies are interdependent and always
appear together.

In the case of $N=1$ the mass formula~(\ref{Etotal}) defines the mass
of an elementary particle, which we denote by $m$.
It will be shown below that for stable bound state solutions of
(\ref{Hartreeeq}-\ref{Hartreepoteq}) we have $m\leq m_0$ and
hence the ground states of particles are bound states and not the
continuum states for which $m=m_0$. It is therefore not an assumption,
but a necessity, that in this model the particles are described by
self-interaction bound states rather than by plane waves or wave
packets. The main aim of this paper is to solve
(\ref{Hartreeeq}-\ref{Hartreepoteq}) for the ground state and to
determine the  dependence of the total mass $M$ and the size parameter
$r_0$ on the coupling constant $g$, the bare mass $m_0$, the mass of
the exchange particle $\mu_0$ and the number of constituents $N$.

In a situation in which all of the constituent particles are in the same
spherically symmetric state with $\vert E_j\vert\equiv\vert E\vert<m_0$
and $\psi_j\left({\bf r}\right)\equiv\,\psi\left({\bf
r}\right)=\,\psi\left(r\right)/\sqrt{4\pi}$,
the system of equations~(\ref{Hartreeeq}-\ref{Hartreepoteq})
simplifies to
\begin{equation}\left(\;\,\frac{{\rm d}^2}{{\rm d}r^2}+\frac{2}{r}\frac{{\rm d}
}{{\rm d}r}-\,\gamma^2+\phi\left(r\right)\;\right)\,\psi\left(r\right)=0
\label{pmodradschreq}\end{equation}
and
\begin{equation}\left(\;\,\frac{{\rm d}^2}{{\rm d} r^2}+\frac{2}{r}\frac{{\rm
 d}}{{\rm d}r}-\mu_0^2\;\right)\phi\left(r\right)=-\frac{g^2N}{4\pi}\,\left|
\,\psi\left(r\right)\right|^2\,,\label{pmpotradeq}\end{equation}
where $\gamma^2=m_0^2-E^2$. Since we are mainly interested in the
properties of the ground state, which is the lowest energy solution of
(\ref{pmodradschreq}-\ref{pmpotradeq}), we can restrict ourselves to
this situation. For the mass we obtain
\begin{equation}M=N\left(\vert E\vert+\frac{1}{2}\int{\rm d}^3{\bf
r}\;\phi\left(r\right)\,\left|\,\psi\left(
{\bf r}\right)\right|^2\right)\,.\label{pmenergy}\end{equation}

Contrary to the case of plane waves, the particles in the present model
are extended. As a measure of their spatial extension, we define the
size parameter
\begin{equation}r_0=2\vert E\vert\int{\rm d}^3{\bf r}\;r\left|\,\psi\left({\bf
r}\right)\right|^2\,,
\label{sizegst}\end{equation}
where the factor $2\vert E\vert$ stems from the normalization condition
(\ref{normcond}).

The stationary states of the composite particle are described by
Hartree's equations. This follows from the fact that
(\ref{Hartreepoteq}) possesses the well known solution
\begin{equation}\phi\left({\bf r}\right)\equiv2mV\left({\bf r}\right)=
\frac{g^2}{4\pi}\sum_{i=1}^N\int{\rm d}^3{\bf r}'\;\frac{{\rm e}^{-\mu_0
\vert{\bf r}-{\bf r}'\vert}}{\vert{\bf r}-{\bf r}'\vert}\,\left|\,\psi_i\left(
{\bf r}'\right)\right|^2\,,\label{Hartreepot}\end{equation}
which, if inserted in~(\ref{Hartreeeq}), gives the usual Hartree's equations
with the effective potential $V\left({\bf r}\right)$. Notice that the
self-interaction ($i=j$) is included in the Hartree's equations and
therefore the case of an elementary particle, where $N=1$, is a
nontrivial special case of this model. Also notice that in the present
model the Hartree's equations are exact and not approximate as in the
usual case.
The model would be more realistic if the constituent
particles were fermions. Yet, despite this shortcoming, it is useful.
For instance, it was shown in ref.~\cite{Witten} that baryons in the
large $N_c$ limit of QCD are described by nonrelativistic
Hartree's equations
\begin{equation}\left[\nabla^2+2m\epsilon+\varphi\left({\bf r}\right)\right]
\chi=0\,,\label{Witteneq}\end{equation}
where
\begin{equation}\varphi\left(\bf r\right)=2mg'^2\int{\rm d}^3{\bf r}'
\;\frac{\left|\chi\left({\bf r}'\right)\right|^2}
{\left|{\bf r}-{\bf r}'\right|}\,,\label{Wittenpot}\end{equation}
$\epsilon$ is the binding energy, $m$ is the quark mass and $\chi$ is the
nonrelativistic single-quark wave function which is normalized according to
$\int{\rm d}^3{\bf  r}\left|\chi\left({\bf r}\right)\right|^2=1$.
Very briefly the reason for this is the following. The color part of
the baryonic wave function is a singlet, which is antisymmetric, and
hence the rest of the wave function must be symmetric. The color and
spin degrees of freedom can be neglected in the lowest order of
approximation and, at the scale of hadronic bound states, the
linear (confining) part of the potential can also be neglected.
Neglecting the relativistic kinematics and certain many-particle
effects, the resulting equations are the nonrelativistic Hartree's
equations~(\ref{Witteneq}-\ref{Wittenpot}) for $N$ identical quarks.
In this paper we will include the relativistic kinematics and solve
the relativistic Hartree's equations
(\ref{Hartreeeq}-\ref{Hartreepoteq}), since our method works equally
well in this case.

Thus, applied to the quark model, $V\left({\bf r}\right)$ in
(\ref{Hartreepot}) with $\mu_0=0$ can be interpreted as
an effective potential in which dressed constituent quarks are bound.
The quark-quark potential for point-like quarks is
$g^2/\left[4\pi\left(2m\right)^2r\right]$, which is obtained by
substituting $\left|\psi_i\left({\bf r}'\right)\right|^2=\delta\left(
{\bf r}'\right)/2m$ in~(\ref{Hartreepot}). Comparing this potential
with the QCD motivated quark-quark potential for point-like quarks,
$4\alpha_s/3r$, we obtain
\begin{equation}\frac{g^2}{4\pi4m^2}=\frac{4\alpha_s\left(m\right)}{3}\,,
\label{QCDident}\end{equation}
where $\alpha_s\left(m\right)$ is the strong running coupling constant of QCD
taken at the mass $m$ of the constituent quark. Equation~(\ref{QCDident})
can be used to relate the results obtained for the present model
to the large $N_c$ limit of QCD. The relation to the nonrelativistic
Hartree's equations~(\ref{Witteneq}-\ref{Wittenpot}) is determined by
$g'^2=g^2N/4\pi4m\left|E\right|$ and $2m\epsilon=-\gamma^2$. Witten drew
two main conclusions in arriving at Hartree's equations
(\ref{Witteneq}-\ref{Wittenpot}). First, provided that $g'$ does not
depend on $N_c$ (i.e. $g\propto1/\sqrt{N_c}$), the baryon masses
increase linearly with $N_c$. This is an immediate consequence of
(\ref{pmenergy}). Second, under the same provision, the size and shape
of the baryons do not depend on $N_c$. This follows from
(\ref{sizegst}) for the size and from
(\ref{pmodradschreq}-\ref{pmpotradeq}) for the shape. However, to
derive further conclusions concerning, for instance, the mass spectrum
or the dependence of the mass on the coupling constant, requires the
solution of the system ~(\ref{pmodradschreq}-\ref{pmpotradeq}).
We will take up these issues here in a more general context where
$\mu_0$ is arbitrary. Besides the connection to QCD, the model is
useful on its own right since it is self-consistent, the dimension is
natural ($3+1$) and the interaction is realistic (exchange of
particles) and includes the self-interaction. Therefore, it allows us
to study the internal dynamical properties of particles such as the
relation between the bare mass $m_0$ and the physical mass $m$, or the
dependence of $m$ on the coupling constant $g$.

The main results of this paper are as follows.

\underline{\em Mass.}$\quad$ For $\mu_0=0$ there is a doublet of spherically
symmetric solutions in the $N=1$ sector, which corresponds to a
doublet of elementary particles with masses $m$ and $m^\ast$, where
$m\leq m^\ast$. Consequently the composite particle with $N$ constituents
possesses $N+1$ states with spherical symmetry, which can be classified
according to how many of the constituents are in the excited state. The
lowest state of the composite particle on the energy scale is the ground
state with the mass $M_N$ given by
\[M_N=Nm_0\,\frac{\sqrt{2}}{3}\left(\,\sqrt{1+\sqrt{1-\left(\frac{g^2N}{4\pi
m_0^2\alpha_0}\right)^2}}+\,\frac{4\pi m_0^2\alpha_0}{g^2N}\,\sqrt{1-\sqrt{1-
\left(\frac{g^2N}{4\pi m_0^2\alpha_0}\right)^2}}\;\right)\,,\]
where $\alpha_0$ is a numerical constant.

For $\mu_0\neq0$ there is a triplet of spherically symmetric solutions
in the $N=1$ sector, which corresponds to a triplet of elementary
particles with masses $m$, $m^\ast$ and $m^{\ast\ast}$, where $m\leq
m^\ast$ and $m\leq m^{\ast\ast}$. Consequently the composite particle
with $N$ constituents possesses now $N\left(N+1\right)/2+N+1$ states
with spherical symmetry, which again can be classified according to
how many of the constituents occupy one or the other excited states.
The dependence of $M/Nm_0$ on $g^2N/4\pi m_0^2$ for various fixed
values of $\mu_0/m_0$ is illustrated in figure~\ref{MassN}a. In this
figure the ground state is plotted with a solid line, the excited
state, having all of the constituents residing in the state~$\ast$, is
plotted with a dashed line and the excited state, having all of the
constituents residing in the state $\ast\ast$, is plotted  with a
dotted-dashed line. Notice that, for each value of $\mu_0/m_0$, the
mass $M$ of the ground state acquires a local minimum at the maximally
allowed value of $g^2/4\pi m_0^2$ (dotted vertical line), which
becomes the absolute minimum $M/Nm_0=2\sqrt{2}/3$ in the case
$\mu_0=0$ (dotted horizontal line). The appearance of such well
pronounced minima of the mass may be a phenomenon which is more
general than one restricted to the present model (see
ref.~\cite{nlschr}, for instance) and hence may have some deeper
significance (some speculations were discussed in ref.~\cite{Pthree}).

\underline{\em Size.}$\quad$ For $\mu_0=0$ the size parameter of the
ground state is given by
\[r_0=\frac{\sqrt{2}\delta_0}{ m_0}\frac{4\pi m_0^2}{g^2N}\sqrt{1+\sqrt{1-
\left(\frac{g^2N}{4\pi m_0^2\alpha_0}\right)^2}}\,,\]
where $\delta_0$ and $\alpha_0$ are certain numerical constants.
For the ground state \mbox{$r_0\rightarrow\infty$}  as
\mbox{$g/m_0\rightarrow0$}. This singularity disappears when
$\mu_0\neq0$. For the case of minimal mass the size parameter of the
ground state becomes minimized too and is given by
\(r_0={\sqrt{2}\delta_0}/\alpha_0m_0\).

\underline{\em Stability.}$\quad$ The stability properties are
investigated according to three independent criteria. The first one is
based on the mass defect \(\Delta M=Nm-M\), where $M$ is the mass of the
composite particle and $m$ denotes the mass of its free constituents.
We found that the ground state is stable against disintegration,
i.e.~$\Delta M>0$.

If the mass of a particle is larger than its bare mass, then such a
particle is unstable. The reason is that, in this case, some of its
continuum states (for instance the states in~(\ref{plwavesol})) are
energetically preferable to the bound state of the particle. According
to this second criterion, in the case $\mu_0=0$ the ground state is
stable. For $\mu_0\neq0$ the ground state is unstable for smaller
values of the coupling constant and stable for larger values (see
figure~\ref{MassN}a). For $\mu_0>0.36m_0$ the ground state becomes
unstable for all possible values of the coupling constant.

The third stability criterion is based on $\delta^2 E_{\rm total}>0$,
which means that a bound state solution is locally stable if it
corresponds to a local minimum of the total energy ($\delta E_{\rm
total}=0$). This criterion was shown by Rosen~\cite{Rosen} to be
necessary and sufficient for a dynamical stability in the sense of
Liapunov. In the present paper we prove that for $\mu_0<\sqrt{2}m_0$
the ground state is locally stable. Moreover, it is proved that our
model is the only option with locally stable bound states among the
class of theories based on lagrangians of the form~(\ref{lagr}), where
$p$ and $q$ are arbitrary real positive constants.

\underline{\em Existence condition.}$\quad$ A short range interaction cannot
produce a bound state unless the strength of the interaction is
sufficiently large, i.e.~the coupling constant is larger
than a certain limit. If the system is relativistic then, in addition
to this constraint, the coupling constant must be smaller than a
certain limit. For the present model this existence condition is given
by
\[\frac{1}{ N}\frac{g_0^2}{4\pi m_0^2}\leq\frac{g^2}{4\pi m_0^2}\leq\frac{1}
{ N}\frac{g_1^2}{4\pi m_0^2}\,,\]
where $g_0$ and $g_1$ are certain functions of $\mu_0/m_0$ and are
independent of $N$.

\underline{\em Wave functions.}$\quad\!$ Equations~(\ref{pmodradschreq})
and~(\ref{pmpotradeq}) are solved using an integral transform method.
The resulting wave functions are plotted in figure~\ref{wfplot}a and
\ref{wfplot}b for various values of the parameter
$\sigma=\mu_0/2\gamma$, where $\gamma^2=m_0^2-E^2\geq0$. More
precisely, the scaled functions $\vert E\vert^{1/2}\,\psi\left(r\right)/
\gamma^{3/2}$and $\phi\left(r\right)/\gamma^2$ are plotted, since in
this form the wave functions are both dimensionless and depend
on the dimensionless variables $\gamma r$ and $\sigma$ only. Notice that
$\phi\left(r\right)>0$ for all $r$. There are no bound state solutions
for the case $\phi\left(r\right)<0$.

\section{Integral transform}\label{Integral}

In ref.~\cite{sol} we discussed a method for solving the Schr\"odinger
equation, which is most suitable if at large distances the potential is
proportional to an exponential function. It follows from~(\ref{pmpotradeq})
that this is the case in the present situation and hence, according
to ref.~\cite{sol} the integral transforms appropriate for
(\ref{pmodradschreq}) and~(\ref{pmpotradeq}) are
\begin{equation}\psi\left(r\right)=a_0\,{\rm e}^{-\gamma r}\int_0^\infty{\rm d}
\mu\;{\rm e}^{-\mu r}\varrho\left(\mu\right)\label{wfbstpmod}\end{equation}
and
\begin{equation}\phi\left(r\right)=\,b_0\,{\rm e}^{-\mu_0r}\int_0^\infty{\rm d}
\mu\;{\rm e}^{-\mu r}{\cal V}\left(\mu\right)\,,\label{Lpotpmod}\end{equation}
where $a_0$ is a normalization constant and $b_0$ is a constant.
Substituting~(\ref{wfbstpmod}) into~(\ref{pmodradschreq}) and
(\ref{Lpotpmod}) into~(\ref{pmpotradeq}), we obtain
\begin{equation}\varrho\left(\mu\right)=1-b_0\,\theta\left(\mu-\mu_0\right)\int_{\mu_0}^\mu
\frac{{\rm
d}\mu'\;}{\mu'\left(\mu'+2\gamma\right)}\,\frac{\partial}{\partial\mu'}
\int_{\mu_0}^{\mu'}{\rm d}\mu''\;{\cal
V}\left(\mu'-\mu''\right)\,\varrho\left(\mu''-\mu_0
\right)\label{inteqpmod}\end{equation}
and
\begin{eqnarray}
\lefteqn{{\cal V}\left(\mu\right)=1-\frac{g^2Na_0^2}{4\pi b_0}\theta\left(\mu+
\mu_0-2\gamma\right)\int_{2\gamma-\mu_0}^\mu\frac{{\rm d}\mu'\;}{\mu'\left(
\mu'+2\mu_0\right)}\,\frac{\partial}{\partial\mu'}}\hspace{6truecm}\nonumber\\
&&\times\int_{2\gamma-\mu_0}^{\mu'}{\rm d}\mu''\,\varrho\left(\mu'-\mu''\right)
\varrho\left(\mu''+\mu_0-2\gamma\right),\label{inteqpotpmod}\end{eqnarray}
where $\theta\left(\mu-a\right)$ is a step-function which vanishes for
$\mu\,<\,a$, equals $1/2$ for $\mu=a$ and is equal to $1$ otherwise and
$a=\mu_0$ or $a=2\gamma-\mu_0$ respectively. In deriving~(\ref{inteqpotpmod})
we have assumed that $\mu_0/2\gamma\leq1$. It will be shown below that
this implies that we are restricting ourselves to the case
$\mu_0\leq2m_0$. Now we introduce the dimensionless variables
\begin{equation}s=\frac{\mu}{2\gamma}\,,\quad\sigma=\frac{\mu_0}{2\gamma}\,,\quad\eta=\,\frac{b_0}{2
\gamma}\,,\quad\zeta=\frac{g^2Na_0^2}{16\pi\gamma^2}\label{dlvarpmod}\end{equation}
and redefine the functions $\varrho$ and ${\cal V}$ in terms of these
variables
\begin{equation}R_{\sigma}\left(\eta,\zeta,s\right)=\,\varrho\left(\mu\right)\,,\quad{\cal V}_\sigma\left(
\eta,\zeta,s\right)=\,{\cal
V}\left(\mu\right)\,.\label{dlvarpmone}\end{equation}
Equations~(\ref{inteqpmod}) and~(\ref{inteqpotpmod}) become
\begin{equation}R_{\sigma}\left(\eta,\zeta,s\right)=1-\eta\,\theta\left(s-\sigma\right)\int_{\sigma}^s
\frac{{\rm d} t\;}{ t\left(t+1\right)}\,\frac{\partial}{\partial
t}\int_\sigma^{t}{\rm d} t'\;
{\cal
V}_{\sigma}\left(\eta,\zeta,t-t'\right)\,R_{\sigma}\left(\eta,\zeta,t'-\sigma\right)
\label{pminteqdl}\end{equation}
and
\begin{eqnarray}
\lefteqn{{\cal V}_\sigma\left(\eta,\zeta,s\right)=1-\frac{\zeta}{\eta}\,
\theta\left(s+\sigma-1\right)\int_{1-\sigma}^s\frac{{\rm d} t}{t\left(t+
2\sigma\right)}\frac{\partial}{\partial t}}\hspace{5.5truecm}\nonumber\\
&&\times\int_{1-\sigma}^t{\rm d}t'R_\sigma\left(\eta,\zeta,t-t'\right)
R_\sigma\left(\eta,\zeta,t'+\sigma-1\right)\label{varVsigma}\end{eqnarray}
respectively. Substituting~(\ref{varVsigma}) in~(\ref{pminteqdl}) and
performing an integration by parts, the dependence on
${\cal V}_\sigma\left(\eta,\zeta,s\right)$
is eliminated and we obtain
\begin{eqnarray}\lefteqn{R_{\sigma}\left(\eta,\zeta,s\right)=1-\eta\,\theta
\left(s-\sigma\right)\int_{\sigma}^s\frac{{\rm d} t\;}{t\left(t+1\right)}
R_{\sigma}\left(\eta,\zeta t-\sigma\right)+\zeta\theta\left(s-1\right)
\int_1^s\frac{{\rm d} t}{t\left(t+1\right)}}\hspace{2truecm}\nonumber\\
&&\times\int_1^t{\rm d} t'\frac{R_{\sigma}\left(\eta,\zeta,t-t'\right)}{ t'^2-
\sigma^2}\,\frac{\partial}{\partial t'}\int_1^{t'}{\rm d}t''\;R_{\sigma}\left(
\eta,\zeta,t'-t''\right)\,R_{\sigma}\left(\eta,\zeta,t''-1\right)\,.
\label{pminteqonedl}\end{eqnarray}
We return now to the question of boundary conditions for~(\ref{Hartreeeq})
and~(\ref{Hartreepoteq}) or, equivalently, for~(\ref{pmodradschreq}) and
(\ref{pmpotradeq}). Since both~(\ref{pmodradschreq}) and~(\ref{pmpotradeq}) are
second
order equations, they must be supplemented by two conditions each. We
first consider the case of $\psi\left(r\right)$ where, as usual, the
conditions are: $\psi\left(\infty\right)=0$ and
$\vert\psi\left(0\right)\vert<\infty$.
The first condition has been taken into account by~(\ref{wfbstpmod}). As for
the second one, in order to translate it to a suitable and practical
condition, we write~(\ref{wfbstpmod}) using~(\ref{dlvarpmod})
and~(\ref{dlvarpmone}) as
\begin{equation}\psi\left(r\right)=2\gamma a_0\,{\rm e}^{-\gamma
r}\int_0^\infty{\rm d} s\;
{\rm e}^{-2\gamma
rs}\,R_\sigma\left(\eta,\zeta,s\right)\label{wfbstpmR}\end{equation}
and replace $\vert\psi\left(0\right)\vert<\infty$ by
\begin{equation}\left|\int_0^{s_0}{\rm d} s\;{\rm e}^{-2\gamma
r_0s}\,R_\sigma\left(\eta,\zeta,s
\right)\right|<\Lambda\,,\label{bstcondR}\end{equation}
where $\Lambda$ is some suitable real number. This condition becomes the
true boundary condition as $s_0\rightarrow\infty$ and then
$r_0\rightarrow0$, but for a finite accuracy result some finite
values of $s_0$ and $r_0$ are sufficient. The order of the limits cannot
be interchanged since $R_\sigma\left(\eta,\zeta,s\right)$ as a function of $s$
does not vanish at infinity (its behaviour in the large $s$ regime can
be described roughly as $s^a\cos\left(bs+c\right)$ with some constants $a$, $b$
and $c$). The conditions for $\phi\left(r\right)$ are the same:
$\phi\left(\infty\right)
=0$ and $\vert\phi\left(0\right)\vert<\infty$. Again, the first condition has
been taken into account by~(\ref{Lpotpmod}), whereas for the second it
follows from
\begin{equation}\phi\left(r\right)=2\gamma b_0\,{\rm
e}^{-\mu_0r}\int_0^\infty{\rm d} s\;{\rm e}^{-2
\gamma rs}\,{\cal
V}_\sigma\left(\eta,\zeta,s\right)\label{wfbstpmV}\end{equation}
that $\vert\phi\left(0\right)\vert<\infty$ can be replaced by
\begin{equation}\left|\int_0^{s_0}{\rm d} s\;{\rm e}^{-2\gamma r_0s}\,{\cal
V}_\sigma\left(\eta,
\zeta,s\right)\right|<\Lambda\label{pmeta}\end{equation}
with some finite values for $r_0$ and $s_0$ depending on the accuracy to
be achieved. Equations~(\ref{bstcondR}) and~(\ref{pmeta}) show that the
constants
$\eta$, $\zeta$ and $\sigma$ are not independent.
\par
Thus, our next task is to solve~(\ref{pminteqonedl}) with parameters $\eta$,
$\zeta$ and $\sigma$ satisfying~(\ref{bstcondR}) and~(\ref{pmeta}). This will
be
achieved in two steps: first we solve~(\ref{pminteqonedl}) regarding $\eta$,
$\zeta$ and $\sigma$ as independent and then for each given $\sigma$
we determine $\eta$ and $\zeta$ satisfying~(\ref{bstcondR}) and~(\ref{pmeta}).
Thus, at this second step $\eta$ and $\zeta$ become functions of
$\sigma$.

\section{Solution of integral equation}\label{Solution}

To obtain the solution of~(\ref{pminteqonedl}) we make the {\em Ansatz}
\begin{eqnarray}\lefteqn{R_\sigma\left(\eta,\zeta,s\right)=
\sum_{n=0}^{\left[s/\sigma\right]}\sum_{m=0}^{\left[s-n\sigma\right]}\left(
-\eta\right)^n\zeta^m\,\varphi_{nm}\left(s-n\sigma-m,\sigma\right)}
\hspace{1.32truecm}\nonumber\\
&&\equiv\sum_{n=0}^\infty\sum_{m=0}^\infty\left(-\eta\right)^n\zeta^m\,\theta
\left(s-n\sigma-m\right)\,\varphi_{nm}\left(s-n\sigma-m,\sigma\right)\,,
\label{pmAnsR}\end{eqnarray}
where a number in square brackets represents the largest integer which is
smaller than or equal to that number. Notice that, for a fixed
$s<\infty$, the right-hand side of the above equation is a finite sum,
so the question of convergence does not appear. Substituting
(\ref{pmAnsR}) in~(\ref{pminteqonedl}) and comparing equal powers of $\eta$ and
$\zeta$, we obtain the recurrence relation for the functions $\varphi_{nm}$
\begin{eqnarray}
&&\hspace{-1.5truecm}\varphi_{nm}\left(y,\sigma\right)=\int_0^y\frac{{\rm
d}t}{\left(t+n\sigma+m+\frac{1}{2}\right)^2-\frac{1}{4\,}}\Biggl(\varphi_{n-1,
m}\left(t,\sigma\right)\nonumber\\
&&\hspace{-1truecm}+\sum_{k=0}^n\sum_{l=1}^m\int_0^{t}{\rm d}t'\frac{\varphi_{n
-k,m-l}\left(t-t',\sigma\right)}{\left(t'+k\sigma+l\right)^2-\sigma^2}
\sum_{i=0}^{k}\sum_{j=1}^l\left.\frac{\partial}{\partial t'}\int_0^{t'}{\rm d}
t''\varphi_{k-i,l-j}\left(t'-t'',\sigma\right)\varphi_{i,j-1}\left(t'',\sigma
\right)\right),\label{pmrrelphitwo}\end{eqnarray}
where $n=0,1,\ldots,$ $m=0,1,\ldots$ and
\begin{equation}\varphi_{-1,m}\left(y,\sigma\right)=0\,,\quad\varphi_{00}
\left(y,\sigma\right)=1\,,\label{pmrrelphizero}\end{equation}
which guarantees that the {\em Ansatz} is in fact a solution. Here
and in what follows we adopted the convention that if the upper bound of
a sum is less than the lower bound then the contribution of the sum is
zero. Notice that for fixed $n$ and $m$ the evaluation of the right-hand
side does not require knowledge of the left-hand side. Also notice
that the {\em Ansatz} has been chosen so that the recurrence relation
does not depend on the parameters $\eta$ and $\zeta$. Therefore, once
the $\varphi_{nm}$ functions are calculated,~(\ref{pminteqonedl}) is solved
for any $\eta$ and $\zeta$ and we can use this solution to find the
values of $\eta$ and $\zeta$ satisfying the boundary conditions
(\ref{bstcondR}) and~(\ref{pmeta}). Equation~(\ref{pmAnsR}) combined with the
above
recurrence relation constitutes the exact analytical solution of
(\ref{pminteqonedl}).\par
The recurrence relation~(\ref{pmrrelphitwo}) can be iterated, so that finally
no $\varphi_{nm}$ functions will appear on the right-hand side.
For instance,
\begin{equation}\varphi_{10}\left(y,\sigma\right)=\,\ln\,\frac{\left(\sigma+1\right)\left(y+\sigma\right)}{
\sigma\left(y+\sigma+1\right)}\,,\quad\varphi_{01}\left(y,\sigma\right)=\frac{1}{2\sigma}
\int_0^y\frac{{\rm d}
t}{\left(t+1\right)\left(t+2\right)}\,\ln\,\frac{\left(t+1-\sigma\right)\left(1+\sigma\right)}{\left(t+1+\sigma\right)\left(1-\sigma\right)}\,.\label{phionezero}\end{equation}
However, for higher $n$ and $m$ one obtains multiple-integral
representations of $\varphi_{nm}$ which are not useful for our purposes.
Since one cannot express the functions $\varphi_{nm}$ in
terms of elementary or special functions, one has to devise a numerical
algorithm for their evaluation. A suitable algorithm is given in the
appendix. As a by-product of this algorithm, we also obtain a function
$f_{nm}\left(s,\sigma\right)$, which is very useful since
${\cal V}_\sigma\left(\eta,\zeta,s\right)$ is related to it in a simple way:
\begin{equation}{\cal V}_\sigma\left(\eta,\zeta,s\right)=1-\frac{1}{\eta}
\sum_{n=0}^\infty\sum_{m=1}^\infty\left(-\eta\right)^n\zeta^m\,\theta\left(s+
\sigma-n\sigma-m\right)\,F_{nm}\left(s+\sigma-n\sigma-m,\sigma\right)\,,
\label{vVf}\end{equation}
where
\begin{equation}F_{nm}\left(y,\sigma\right)=\int_0^y{\rm d}t\;f_{nm}\left(t,
\sigma\right)\,.\label{Fdef}\end{equation}
Notice that, for a fixed $s<\infty$, the right-hand side
of~(\ref{vVf}) is a finite sum.

Once the functions $\varphi_{nm}\left(y,\sigma\right)$ and
$F_{nm}\left(y,\sigma\right)$ are calculated we can use them to calculate the
function $R_\sigma\left(\eta,\zeta,s\right)$ according to~(\ref{pmAnsR}) and
the
function ${\cal V}_\sigma\left(\eta,\zeta,s\right)$ according to
(\ref{vVf}), and to determine the values of $\eta$ and $\zeta$ for
which the boundary conditions~(\ref{bstcondR}) and~(\ref{pmeta}) are
fulfilled. A typical result is illustrated in figure~\ref{RVplot}.

\section{Wave functions and binding energies}\label{Wave}

To determine the wave function $\psi\left(r\right)$ of a constituent we have
to determine the normalization constant $a_0$. Using the normalization
condition~(\ref{normcond}) and equation~(\ref{wfbstpmR}), we obtain
\begin{equation}a_0^2=\frac{2\gamma\zeta}{\alpha\vert E\vert}\,,\label{azero}
\end{equation}
where
\begin{equation}\alpha=16\zeta\int_0^\infty{\rm
d}\xi\;\xi^2\,\Psi_\sigma^2\left(\xi\right)\,,\quad\Psi_\sigma\left(\xi\right)
={\rm e}^{-\xi}\int_0^\infty{\rm d} s\;{\rm e}^{-2\xi s}\,R_\sigma\left(
\eta,\zeta,s\right)\,.\label{alphadef}\end{equation}
Notice that $\sigma$ is the only parameter upon which $\alpha$ depends.
Using~(\ref{alphadef}) and the function $R_\sigma\left(\eta,\zeta,s\right)$
calculated
above, $\alpha$ can be determined. A useful fit is
\begin{equation}\alpha=\alpha_0+\alpha_1\sigma+\alpha_2\sigma^2\,,\quad\sigma\leq1\,,
\label{alphafunc}\end{equation}
where $\alpha_0=3.52(2)$, $\alpha_1=10.9(2)$ and $\alpha_2=3.82(9)$.
Using~(\ref{azero}) and
(\ref{wfbstpmR}) we can write the wave function in a dimensionless form:
\begin{equation}\frac{\vert
E\vert^{1/2}}{\gamma^{3/2}}\,\psi\left(r\right)=2\sqrt\frac{2\zeta}{\alpha}
\,{\rm e}^{-\gamma r}\int_0^\infty{\rm d} s\;{\rm e}^{-2\gamma
rs}\,R_\sigma\left(\eta,
\zeta,s\right)\,.\label{wfdless}\end{equation}
Notice that the right-hand side of~(\ref{wfdless}) is a function of the
dimensionless quantities $\gamma r$ and $\sigma$ only. Similarly,
using~(\ref{wfbstpmV}) and~(\ref{dlvarpmod}) the potential function
$\phi\left(r\right)$
also can be written in the dimensionless form
\begin{equation}\frac{1}{\gamma^2}\,\phi\left(r\right)=4\eta\,{\rm
e}^{-2\sigma\gamma r}\int_0^\infty
{\rm d} s\;{\rm e}^{-2\gamma rs}\,{\cal
V}_\sigma\left(\eta,\zeta,s\right)\,.\label{Vdless}\end{equation}
In figures \ref{wfplot}a and \ref{wfplot}b we plotted the wave function
and the potential function according to~(\ref{wfdless}) and
(\ref{Vdless}) respectively for various values of $\sigma$.

We define the binding energy $E-m_0$ of a constituent particle as
the difference between its constituent energy $E$ and its bare mass
$m_0$. To determine the constituent energy $E$ we substitute~(\ref{azero})
into the expression for $\zeta$ provided by~(\ref{dlvarpmod}) and using
$\gamma^2=m_0^2-E^2$ obtain
\begin{equation}\vert E\vert\sqrt{m_0^2-E^2}=\frac{1}{2}\,\frac{g^2N}{4\pi
\alpha}\,.\label{benergy}\end{equation}
Solving for $\vert E\vert$ we obtain two solutions:
\begin{equation}\varepsilon^\ast=\frac{m_0}{\sqrt{2}}\,\sqrt{1-\sqrt{1-\left(
\frac{g^2N}{4\pi m_0^2\alpha}\right)^2}}\label{constexen}\end{equation}
and
\begin{equation}\varepsilon=\frac{m_0}{\sqrt{2}}\,\sqrt{1+\sqrt{1-\left(
\frac{g^2N}{4\pi m_0^2\alpha}\right)^2}}\,,\label{consten}\end{equation}
which are real if $g^2N/4\pi m_0^2\alpha\leq1$. The first root is the
constituent energy of each constituent in the excited state, while the
second root corresponds partly to the ground state and partly to the
excited state, depending on the value of $\sigma$. Equations
(\ref{constexen}) and~(\ref{consten}) imply that
\begin{equation}0<\frac{\varepsilon^\ast}{ m_0}\le\frac{1}{\sqrt{2}}
\label{exenineq}\end{equation}
and
\begin{equation}\frac{1}{\sqrt{2}}\le\frac{\varepsilon}{ m_0}<1\,.
\label{enineq}\end{equation}
Notice that $\varepsilon^\ast$ is not larger than $\varepsilon$, and
nevertheless $\varepsilon^\ast$ is assigned entirely to the excited
state. The reason is that, as will be shown below, the total energy
(the mass) corresponding to the bound state with constituent energy
$\varepsilon^\ast$ is always larger than in the case of a bound state
with constituent energy $\varepsilon$.

In the case when $\mu_0=0$, the function $\alpha=\alpha_0=3.52(2)$ is a
pure numerical constant and hence~(\ref{constexen}) and~(\ref{consten}) are
explicit formulae. Otherwise, $\alpha$ is a function of $\sigma$ which
itself is a function of both parameters $\mu_0/m_0$ and $g$. We will not
need the explicit form of this function, since $\sigma$ will serve only
as a parameterization variable at fixed $\mu_0/m_0$. The allowed range of
$\sigma$ follows from
\begin{equation}\sigma=\frac{1}{2}\frac{\mu_0}{m_0}\,\frac{1}{\sqrt{1-\left(
\frac{E}{ m_0}\right)^2}}\,,\label{sigmafunc}\end{equation}
which is obtained from the definition $\sigma=\mu_0/2\gamma$ and from
$\gamma^2=m_0^2-E^2$. Substituting~(\ref{exenineq}) and~(\ref{enineq})
in~(\ref{sigmafunc}), we obtain respectively
\begin{equation}\frac{1}{2}\frac{\mu_0}{
m_0}<\sigma\leq\frac{1}{\sqrt{2}}\frac{\mu_0}{ m_0}
\label{sigmaineqex}\end{equation}
and
\begin{equation}\frac{1}{\sqrt{2}}\frac{\mu_0}{
m_0}\leq\sigma<\infty\,.\label{sigmaineq}\end{equation}
The coupling constant $g^2N/4\pi m_0^2$, for instance, is parameterized
by $\sigma$ at fixed $\mu_0/m_0$ via
\begin{equation}\frac{g^2N}{4\pi m_0^2}=\frac{\alpha\mu_0}{\sigma
m_0}\sqrt{1-\frac{1}{4}
\left(\frac{\mu_0}{\sigma m_0}\right)^2}\,.\label{gexpl}\end{equation}
Equation~(\ref{gexpl}) is obtained if one combines~(\ref{benergy}) and
(\ref{sigmafunc}) to eliminate $E$. Using~(\ref{gexpl}) we can convert
the dependence on $\sigma$ in~(\ref{constexen}) and~(\ref{consten}) to
a dependence on $g^2N/4\pi m_0^2$ and $\mu_0/m_0$. The result,
$\varepsilon/m_0$ and $\varepsilon^\ast/m_0$ as functions of
$g^2N/4\pi m_0^2$ and $\mu_0/m_0$, is plotted in figure~\ref{eofgn}.
The dotted line indicates the value $1/\sqrt{2}$ which is the boundary
value deviding the two functions.

\section{Existence condition}\label{Existence}

Equations~(\ref{constexen}) and~(\ref{consten}) imply that
\begin{equation}0<\frac{g^2}{4\pi m_0^2}\leq\frac{\alpha}{ N}\,,
\label{excondzero}\end{equation}
since otherwise the constituent energy becomes a complex number,
in which case the bound state does not exist.
In the case when $\mu_0=0$, $\alpha=\alpha_0=3.52(2)$ is a pure
numerical constant and hence~(\ref{excondzero}) is an explicit necessary
and sufficient condition for the existence of bound states.

In the case when $\mu_0\neq0$~(\ref{excondzero}) is not suitable as an
explicit existence condition, since in this case $\alpha$ is not a
pure number but rather a function of $\sigma$ and thus of $g^2N/4\pi
m_0^2$ and $\mu_0/m_0$. To obtain the existence condition for the case
when $\mu_0\neq0$, we write~(\ref{gexpl}) as
\begin{equation}\frac{g^2}{4\pi m_0^2}=\frac{\alpha}{ N}\,\frac{\mu_0}
{\sigma m_0}\sqrt{1-\frac{1}{4}\left(\frac{\mu_0}{\sigma m_0}\right)^2}\leq
\frac{\alpha}{ N}\,,\label{gexpla}\end{equation}
which is the parametrization of the coupling constant in terms of
$\sigma$ at fixed $\mu_0/m_0$ and $N$. The inequality on the right-hand
side of~(\ref{gexpla}) becomes an equality for $\sigma=\mu_0/\sqrt{2}m_0$.
Since $\sigma$ varies in a definite range given in~(\ref{sigmaineqex}) and
(\ref{sigmaineq}), equation~(\ref{gexpla}) determines the allowed range of the
coupling constant, in which a bound state solution is possible.
In figure~\ref{gNofsig} we plot $g^2N/4\pi m_0^2$ as a function of
$\sigma$ for a few typical values of $\mu_0/m_0$ (solid lines). Notice
that $g^2N/4\pi m_0^2$ possesses a local maximum in the region of smaller
$\sigma$ and a local minimum in the region of larger $\sigma$.
{}From~(\ref{gexpla}) and~(\ref{alphafunc}) we obtain
\begin{equation}\sigma_1=\frac{1}{\sqrt{2}}\frac{\mu_0}{m_0}+\frac{\alpha_1}
{8\alpha_0}\left(\frac{\mu_0}{m_0}\right)^2+O\left(\left(\frac{\mu_0}{m_0}
\right)^2\right)\label{sigmazero}\end{equation}
for the location $\sigma=\sigma_1$ of the local maximum and
\begin{equation}\sigma_0=\sqrt\frac{\alpha_0}{\alpha_2}+O\left(\left(
\frac{\mu_0}{ m_0}\right)^2\right)\label{sigmaone}\end{equation}
for the location $\sigma=\sigma_0$ of the local minimum.
The respective accuracies of~(\ref{sigmazero}) and~(\ref{sigmaone})
are sufficient for all practical purposes. This can be judged by
looking at figure~\ref{gNofsig}, where $\sigma_1$ from
(\ref{sigmazero}) is plotted as a vertical dotted line and
$\sigma_0=\sqrt{\alpha_0/\alpha_2}$ is plotted as a vertical dashed
line. The maximum and the minimum split the allowed range of $g^2/4\pi
m_0$ into three regions: one to the left of the maximal value
\begin{equation}0<\frac{g^2}{4\pi m_0^2}\leq\frac{1}{N}\frac{g_1^2}{4\pi m_0^2
}\,,\quad\frac{1}{2}\frac{\mu_0}{m_0}<\sigma\leq\sigma_1\,,\label{glimex}
\end{equation}
one between the maximal value and the minimal value
\begin{equation}\frac{1}{ N}\frac{g_0^2}{4\pi m_0^2}\leq\frac{g^2}{4\pi m_0^2}
\leq\frac{1}{ N}\frac{g_1^2}{4\pi m_0^2}\,,\quad\sigma_1\leq\sigma\leq\sigma_0
\,,\label{glim}\end{equation}
and one to the right of the minimal value
\begin{equation}\frac{1}{ N}\frac{g_0^2}{4\pi m_0^2}\leq\frac{g^2}{4\pi m_0^2}
\,,\quad\sigma_0\leq\sigma\,,\label{glimone}\end{equation}
where
\begin{equation}\frac{g_1^2}{4\pi m_0^2}=\alpha\left(\sigma_1\right)\,\frac{\mu
_0}{\sigma_1m_0}\sqrt{1-\frac{1}{4}\left(\frac{\mu_0}{\sigma_1m_0}\right)^2}=
\alpha_0+\frac{\alpha_1}{\sqrt{2}}\frac{\mu_0}{m_0}+O\left(\left(\frac{\mu_0}
{ m_0}\right)^2\right)\,,\label{gnotexpl}\end{equation}
\begin{equation}\frac{g_0^2}{4\pi m_0^2}=\alpha\left(\sigma_0\right)\,\frac{\mu
_0}{\sigma_0m_0}\sqrt{1-\frac{1}{4}\left(\frac{\mu_0}{\sigma_0m_0}\right)^2}=
\left(2\sqrt{\alpha_0\alpha_2}+\alpha_1\right)\frac{\mu_0}{m_0}+O\left(\left(
\frac{\mu_0}{ m_0}\right)^3\right)\label{goneexpl}\end{equation}
and $\alpha$ is given in~(\ref{alphafunc}). These three branches of
the coupling constant correspond to three different states
(particles), and below it will be shown that~(\ref{glim}) corresponds
to the ground state while~(\ref{glimex}) and~(\ref{glimone}) each
correspond to a different excited state. Hence for the case of
$\mu_0\neq0$ the model predicts a triplet of particles. Notice that in
the case when $\mu_0=0$~(\ref{glim}) becomes
\begin{equation}0<\frac{g^2}{4\pi m_0^2}\leq\frac{\alpha_0}{ N}\,.
\label{glimzero}\end{equation}
If for certain $\mu_0/m_0$ and $N$ it happens that $g^2/4\pi m_0^2$ is
outside of the bounds defined in~(\ref{glim}) then a ground state solution
and a particle associated with it do not exist and there are no stable
particles in that case. Thus~(\ref{glim}) is an explicit condition for the
existence of a ground state and its associated particle. Notice that
(\ref{glim}) is in  qualitative agreement with the nonrelativistic case of
the Yukawa potential~\cite{sol}, where the ground state solution is
possible only if the coupling constant $f^2/4\pi$ is larger than
$\left(1.6798/2\right)\left(\mu_0/m_0\right)$.

In the case when the condition~(\ref{glim}) for the existence of the ground
state is violated while~(\ref{glimex}) is fulfilled, the particle associated
with the ground state does not exist while the excited state does.
However, as will be shown below, the mass of such a state is larger
than its bare mass. Hence this state is unstable and represents an
unstable particle.

\section{Mass and bare mass}\label{Mass}

Instead of calculating the last term in~(\ref{pmenergy}) directly, we can
use the virial theorem (this theorem is proved in section
\ref{Derrick} below) part of which is the relation
\begin{equation}\frac{1}{2}\int{\rm d}^3{\bf
r}\;\phi\left(r\right)\left|\,\psi\left({\bf r}\right)\right|^2=
\frac{m_0}{3}\left(\frac{m_0}{\vert E\vert}-\frac{\vert E\vert}{
m_0}\right)+\frac{1}{3}\frac{\mu_0^2}{
g^2N}\int{\rm d}^3{\bf
r}\;\phi^2\left(r\right)\,.\label{pmvirtheor}\end{equation}
Substituting~(\ref{pmvirtheor}) in~(\ref{pmenergy}) we obtain
\begin{equation}M=Nm_0\left(\frac{2}{3}\frac{\vert E\vert}{
m_0}+\frac{1}{3}\frac{m_0}{\vert E\vert}+\frac{1}{3}
\frac{\mu_0^2}{ m_0g^2N}\int{\rm d}^3{\bf r}\;\phi^2\left(r\right)\right)>0\,.
\label{pmenergyone}\end{equation}
Notice that the total energy~(\ref{pmenergyone}) is positive definite.
\par
We first consider the case when $\mu_0=0$. In this case the last term
in~(\ref{pmenergyone}) vanishes and we obtain
\begin{equation}M=Nm_0\frac{2}{3}\left(\frac{\vert E\vert}{
m_0}+\frac{1}{2}\frac{m_0}{\vert E\vert}\right)\,.
\label{Masseps}\end{equation}
Substitution of~(\ref{consten}) and~(\ref{constexen}) in~(\ref{Masseps}) yields
\begin{equation}M_N=Nm_0\,\frac{\sqrt{2}}{3}\left(\,\sqrt{1+\sqrt{1-\left(\frac{g^2N}{4\pi m_0^2\alpha_0}
\right)^2}}+\,\frac{4\pi m_0^2\alpha_0}{
g^2N}\,\sqrt{1-\sqrt{1-\left(\frac{g^2N}{4\pi m_0^2
\alpha_0}\right)^2}}\;\right)\label{MQCD}\end{equation}
for the ground state and
\begin{equation}M_{NN}=Nm_0\,\frac{\sqrt{2}}{3}\left(\,\sqrt{1-\sqrt{1-\left(\frac{g^2N}{4\pi m_0^2
\alpha_0}\right)^2}}+\,\frac{4\pi m_0^2\alpha_0}{
g^2N}\,\sqrt{1+\sqrt{1-\left(\frac{g^2N
}{4\pi m_0^2\alpha_0}\right)^2}}\;\right)\label{MoneQCD}\end{equation}
for the excited state. The masses of the corresponding constituent
particles are
\begin{equation}m=M_1\,,\quad m^\ast=M_{11}\,.\label{moneQCD}\end{equation}
Thus, for the case of $\mu_0=0$ the model predicts a doublet in the
elementary particle sector ($N=1$) with masses $m,\;m^\ast$. Notice that
(\ref{MQCD}-\ref{moneQCD}) imply that for fixed parameters $g^2N/4\pi m_0^2$
and $Nm_0$
\begin{equation}M_N\leq M_{NN}\,,\quad m\leq m^\ast\label{masscomp}
\end{equation}
despite the fact that according to~(\ref{consten}) and~(\ref{constexen})
$\varepsilon^\ast\leq\varepsilon$. The reason is that, besides the
contribution from the constituent energies (first term in~(\ref{Etotal})),
the mass receives also a contribution from the energy (second term in
(\ref{Etotal})) which must be spent to create the potential in which the
constituents are bound. This contribution compensates the difference
between $\varepsilon^\ast$ and $\varepsilon$ and places $M_{NN}$
above $M_N$. Also notice that while $M_{NN}\rightarrow\infty$ and
$m^\ast\rightarrow\infty$ as $g\rightarrow0$, $M_N$ and $m$ stay finite.
\par
The state with mass $M_{NN}$ is the highest excited state with
spherical symmetry, since we have assumed that all of the constituents
occupy the same state and that there are only two bound states
which each of the constituents can occupy.  Our method, however, can be
extended easily to a situation where one or more constituents occupy the
excited state with the corresponding constituent energy
$\varepsilon^\ast$ while others remain in the other bound state. The
resulting excited states will possess masses between $M_N$ and
$M_{NN}$:
\begin{equation}M_N\leq M_{1N},M_{2N},\cdots\leq
M_{NN}\,.\label{exstmass}\end{equation}
These are all the possible spherically symmetric states, since
otherwise there would be more states than just a doublet in the $N=1$
sector. Hence for the case of $\mu_0=0$, the total number of states with
spherical symmetry is
\begin{equation}N_{\rm states}=N+1\,.\label{Nstmunot}\end{equation}
\par
A further consequence of~(\ref{MQCD}-\ref{moneQCD}) is
\[\Delta M_N=Nm-M_N>0\,,\]
which means that the ground state of the composite particle is stable
against disintegration. Similarly, one obtains
\begin{equation}\Delta
M_{NN}=Nm^\ast-M_{NN}>0\,,\label{Monemonecomp}\end{equation}
which means that the excited state cannot decay by disintegration
with emission of particles of mass $m^\ast$.
\par
Above we discussed the stability of the composite particle from the
point of view of disintegration. We now discuss the stability
from the point of view of a comparison between the bare mass and the
physical mass. This criterion allows us to discuss the stability of even
the elementary particles of the theory. From~(\ref{MQCD}-\ref{moneQCD}) it
follows that
\begin{equation}0<\frac{g^2}{4\pi m_0^2}\leq\frac{\alpha_0}{
N}\quad\Leftrightarrow\quad
M_N<Nm_0\,,\label{confcondone}\end{equation}
for the ground state
and
\begin{equation}0<\frac{g^2}{4\pi m_0^2}\leq\frac{\sqrt{3}}{2}\frac{\alpha_0}{
N}\quad\Leftrightarrow
\quad M_{NN}\geq Nm_0\,,\label{ccond}\end{equation}
\begin{equation}\frac{\sqrt{3}}{2}\frac{\alpha_0}{ N}<\frac{g^2}{4\pi
m_0^2}\leq\frac{\alpha_0}{ N}\quad
\Leftrightarrow\quad M_{NN}<Nm_0\label{ccondone}\end{equation}
for the excited state.
If we want to associate a particle with a self-interaction bound state,
then the result that the physical mass is larger than the bare mass
means that such a particle is unstable. The reason is that in this case
some of its continuum states are energetically preferable to the bound
state. Thus from~(\ref{confcondone}) we conclude that the ground state $M_N$
is always stable if it exists, while the excited state $M_{NN}$ is
unstable for smaller values of the coupling constant $g$ in the case of
(\ref{ccond}), and stable for larger values of the coupling constant in the
case of~(\ref{ccondone}). The phenomenon of instability due to the mass being
larger than the bare mass is not new.
It was observed, for example, for the ground state in the model of
ref.~\cite{nlschr}, which is based on the nonlinear Schr\"odinger equation
in 3-D. In that model one can give a formal proof (see ref.~\cite{Blowup}
for the details) that the ground state is unstable in the sense that its
wave function suffers a collapse; i.e., being initially smooth it
develops a singularity within a finite period of time when subjected to
a small perturbation. This means that such a state cannot be associated
with a stable free particle.
\par
In the case of~(\ref{ccondone}) the composite particle can exist as
a stable (with respect to the continuum states) particle with the
mass $M_N$ or $M_{NN}$. For the mass difference we have
\begin{equation}0\leq M_{NN}-M_N<\frac{3\sqrt{2}-4}{4}M_N\,,\label{massdiff}
\end{equation}
which amounts to less than $6$ \%.

We now discuss the case when $\mu_0\neq0$. Using~(\ref{Lpotpmod}),
(\ref{dlvarpmod}),~(\ref{benergy}) and~(\ref{sigmafunc}), we obtain
\begin{equation}\frac{1}{2}\frac{\mu_0^2}{ m_0g^2N}\int{\rm d}^3{\bf r}\,
\phi^2\left(r\right)=\left(\frac{m_0}{\vert E\vert}-\frac{\vert E\vert}{ m_0}
\right)\beta\,,\label{phibeta}\end{equation}
where
\begin{equation}\beta=\frac{16\sigma^2\eta^2}{\alpha}\int_0^\infty{\rm d}\xi\;
\xi^2\phi_\sigma^2\left(\xi\right)\,,\quad\phi_\sigma\left(\xi\right)=
{\rm e}^{-2\sigma\xi}\int_0^\infty{\rm d} s\;{\rm e}^{-2\xi s}\,{\cal V}_\sigma
\left(\eta,\zeta,s\right)\label{betadef}\end{equation}
and ${\cal V}_\sigma\left(\eta,\zeta,s\right)$ is defined in~(\ref{varVsigma}).
Notice that $\beta$ is a function of $\sigma$ only. To get some idea of the
behavior of $\beta$ the following fit is useful
\begin{equation}\beta=\beta_0\sigma\,\frac{1+\beta_1\sigma}{1+\beta_2\sigma}\,,\quad\sigma
\leq1\,,\label{betaappr}\end{equation}
where $\beta_0=0.834$, $\beta_1=0.972$ and $\beta_2=2.93$. The accuracy
of~(\ref{betaappr}) is about $1$\%. Using~(\ref{phibeta}) we can rewrite
(\ref{pmenergyone}) as
\begin{equation}M=Nm_0\frac{2}{3}\left(\left(1-\beta\right)\frac{\vert E\vert}{
m_0}+\frac{1}{2}\left(1+2\beta\right)\frac{m_0
}{\vert E\vert}\right)\,.\label{pmetwo}\end{equation}
Substituting~(\ref{constexen}) and~(\ref{consten}) in~(\ref{pmetwo}) we can
obtain
similar mass formulae to~(\ref{MQCD}-\ref{moneQCD}). More convenient formulae
are obtained if, using~(\ref{gexpla}), we eliminate $g^2N/4\pi m_0^2$ from
(\ref{constexen}) and~(\ref{consten}) and substitute the result
in~(\ref{pmetwo}).
The resulting mass formulae are
\begin{equation}M_N=Nm_0\,\Xi\left(\sigma,\frac{\mu_0}{
m_0}\right)\,,\quad\sigma_0\geq\sigma\geq
\sigma_1=\frac{1}{\sqrt{2}}\frac{\mu_0}{
m_0}+\frac{\alpha_1}{8\alpha_0}\left(\frac{\mu_0}{ m_0}
\right)^2+O\left(\left(\frac{\mu_0}{
m_0}\right)^2\right)\,,\label{pmethree}\end{equation}
\begin{equation}M_{0NN}=Nm_0\,\Xi\left(\sigma,\frac{\mu_0}{
m_0}\right)\,,\quad\sigma_1\geq\sigma>
\frac{1}{2}\frac{\mu_0}{ m_0}\,,\hskip5.8truecm\label{pmefour}\end{equation}
and
\begin{equation}M_{N0N}=Nm_0\,\Xi\left(\sigma,\frac{\mu_0}{
m_0}\right)\,,\quad\infty>\sigma>\sigma_0
=\sqrt\frac{\alpha_0}{\alpha_2}+O\left(\left(\frac{\mu_0}{
m_0}\right)^2\right)\,,\hskip2.3truecm
\label{pmefive}\end{equation}
where
\begin{equation}\Xi\left(\sigma,\frac{\mu_0}{
m_0}\right)=\frac{1-\frac{1-\beta}{6}\left(\frac{\mu_0}{\sigma m_0}\right)^2}{1
-\frac{1}{4}\left(\frac{\mu_0}{\sigma
m_0}\right)^2}\sqrt{1-\frac{1}{4}\left(\frac{\mu_0}{\sigma m_0}\right)^2}
\equiv\left(1-\frac{1-\beta}{6}\left(\frac{\mu_0}{\sigma
m_0}\right)^2\right)\frac{\mu_0}{\sigma m_0}\frac{4\pi
m_0^2\alpha}{ g^2N}\,.\label{Msigmum}\end{equation}
The advantage of~(\ref{pmethree}-\ref{pmefive}) is that, for a fixed
$\mu_0/m_0$ and $Nm_0$, they depend solely on $\sigma$.
The dependence of $M_N$, $M_{0NN}$ and $M_{N0N}$ on $\sigma$ for
various fixed values of $\mu_0/m_0$ is illustrated in figure~\ref{MassN}b.
In this figure $M_N/Nm_0$ is plotted with the solid line,
$M_{0NN}/Nm_0$ with the dashed line and $M_{N0N}/Nm_0$ with
the dotted-dashed line. The dependence on $\sigma$ can be converted
to the dependence on $g^2N/4\pi m_0^2$ via~(\ref{gexpla})
\begin{equation}\frac{g^2N}{4\pi m_0^2}=\alpha\,\frac{\mu_0}{\sigma
m_0}\sqrt{1-\frac{1}{4}\left(\frac{\mu_0}{\sigma
m_0}\right)^2}\,.\end{equation}
The result is illustrated in figure~\ref{MassN}a. In both figures the
dotted vertical lines correspond to the maximal values of $g^2N/4\pi
m_0^2$ according to~(\ref{gexpla}) ($\sigma=\sigma_1$), while the dotted
horizontal line corresponds to the absolute minimum of the mass
$M/Nm_0=2\sqrt{2}/3$, which occurs at $\mu_0=0$. Hence we observe that
the mass $M_N$ of the ground state acquires a minimum value at the
maximal value of $g^2/4\pi m_0^2$. We make a similar observation
with respect to the maximal value of the mass $M_N$: this occurs
at the minimal value of $g^2/4\pi m_0^2$ according
to~(\ref{gexpla}) ($\sigma=\sigma_0$). Notice that, except for the case when
$\mu_0/m_0=0$, part of $M_N$ including the maximum resides within the
continuum, i.e. above $Nm_0$, signaling an unstable state. This part
becomes larger with increasing $\mu_0/m_0$ and starting with $\mu_0/m_0=
0.36$ all of the mass $M_N$ resides within the continuum. This means that
for a stable ground state the mass of the exchange particle cannot be
too large: $\mu_0<0.36m_0$.

In the case of elementary particles of the model ($N=1$) we have a
triplet with the corresponding masses
\begin{equation}m=M_1\,,\quad m^\ast=M_{011}\,,\quad m^{\ast\ast}=M_{101}\,,
\label{mmasses}\end{equation}
which are obtained by putting $N=1$ in~(\ref{pmethree}-\ref{pmefive}).
This fact is in contrast to the $\mu_0=0$ case where there was a
doublet, and is a clear indication that the mass is not an analytic
function of $\mu_0$ at $\mu_0=0$. The mass is also nonanalytic at
$g=g_1$, where it has a minimum, and at $g=g_0$ where it has a local
maximum (see equations~(\ref{gnotexpl}) and~(\ref{goneexpl})).
Otherwise it is an analytic function of $g^2N/4\pi m_0^2$ and
$\mu_0/m_0$.

The appearance of a triplet with different masses and different
constituent energies at the same values of $g^2N/4\pi m_0^2$ and
 $\mu_0/m_0$ indicates that there are other states with spherical
symmetry, which can be labelled by the number of constituents in each of
the two excited states. The lowest mass state $M_N$ is the ground
state, where none of the constituents reside in an excited state. Then
there are states $M_{k,n-k,N}$ where $n$ constituents reside in an
excited state, $k$ of them in one of the excited states and $n-k$ in
the other one. Since all these states are spherically symmetric, our
method can be easily extended to include them. Basically, the only
significant change in this case is the tripling of the basic equations,
so that we would have six coupled equations instead of two. Since in the
present case there are no other states beyond the triplet, it is fairly
clear that these are all spherically symmetric states of the model and
that their total number is
\begin{equation}N_{\rm states}=\frac{N\left(N+1\right)}{2}+N+1\,.
\label{Nofstates}\end{equation}
Thus in this model there are $3$ elementary particles (triplet), $6$
composite particles with $2$ constituents (sextet), $10$ composite
particles with $3$ constituents (decuplet) etc.
\par
One of the most basic properties of the masses as functions of
$g^2N/4\pi m_0^2$ at fixed $\mu_0/m_0$ is
\begin{equation}\frac{g'^2N}{4\pi m_0^2}\leq\frac{g^2N}{4\pi
m_0^2}\;\Rightarrow\;\frac{M'}{ Nm_0}\geq
\frac{M}{ Nm_0}\,,\label{Mmonot}\end{equation}
where $M'$ and $M$ are any of the masses
$M_N$, $M_{N0N}$, $M_{0NN}$, $m$, $m^\ast$ and $m^{\ast\ast}$ evaluated
at $g'^2N/4\pi m_0^2$ and $g^2N/4\pi m_0^2$ respectively. An immediate
implication of this result is that
\begin{equation}\Delta M_N=Nm-M_N>0\,,\label{Mmcomp}\end{equation}
\begin{equation}\Delta M_{0NN}=Nm^\ast-M_{0NN}>0\,,
\label{MNnotcomp}\end{equation}
\begin{equation}\Delta M_{N0N}=Nm^{\ast\ast}-M_{N0N}>0\,.
\label{MnotNcomp}\end{equation}
Thus none of the composite particles can decay by disintegration and
in particular the ground state is absolutely stable whenever its mass is
below the bare mass. In the case when any of the masses is larger than
the corresponding bare mass, for instance $M_{N0N}\geq Nm_0$ always,
the particle with that mass is unstable as discussed above.

One of the striking features of the mass (total energy) as a function of
the coupling constant $g^2/4\pi m_0^2$ is the appearance of a strongly
pronounced unique minimum at each fixed value of $\mu_0/m_0$ and $N$
(see figure~\ref{MassN}a). These minimal values of the mass and the
corresponding coupling constants can be determined from
(\ref{pmethree}),~(\ref{gexpla}) and~(\ref{sigmazero}) and, to first
order in $\mu_0/m_0$, are given by
\begin{equation}M_N=Nm_0\frac{2\sqrt{2}}{3}\left(1+\frac{\beta_0}{2\sqrt{2}}
\frac{\mu_0}{ m_0}\right)\,,\quad\frac{g^2}{4\pi m_0^2}=\frac{1}{N}\left(
\alpha_0+\frac{\alpha_1}{\sqrt{2}}\frac{\mu_0}{ m_0}\right)\,,\quad
N=1,2,\ldots\,,\label{quantgmu}\end{equation}
where $\alpha_0=3.52(2)$, $\alpha_1=10.9(2)$, $\beta_0=0.834$. The
fact that the physical mass of a particle possesses such an absolute
minimum was noticed earlier in ref.~\cite{nlschr} in connection with
some other model, where it was emphasized that this phenomenon might
be a general one and might have some deep significance (some
speculations are discussed in ref.~\cite{Pthree}).

\section{Size parameter}\label{Size}

The size parameter $r_0$ has been defined in~(\ref{sizegst}). Substituting
(\ref{wfbstpmR}) and~(\ref{azero}) in~(\ref{sizegst}), we obtain
\begin{equation}r_0=\,\frac{\delta}{\gamma}\equiv\frac{\delta}{\sqrt{m_0^2-E^2}}\,,\label{rnot}\end{equation}
where
\begin{equation}\delta=\frac{16\zeta}{\alpha}\int_0^\infty{\rm
d}\xi\;\xi^3\,\Psi_\sigma^2\left(\xi
\right)\,,\label{deldef}\end{equation}
and $\Psi_\sigma\left(\xi\right)$ is defined in~(\ref{alphadef}). As in the
case
of $\alpha$ and $\beta$, the function $\delta$ can also be easily
computed, once the integral equation~(\ref{pminteqonedl}) above is solved. A
useful fit, which is accurate to about $1$\%, is
\begin{equation}\delta=\frac{\delta_0}{\alpha}\sqrt{1+\delta_1\sigma+\delta_2\sigma^2}\,,
\quad\sigma\leq1\,,\label{deltanum}\end{equation}
where $\delta_0=8.4$, $\delta_1=1.35$, $\delta_2=5.75$ and $\alpha$ is
given in~(\ref{alphafunc}).
Thus
\begin{equation}1.3<\delta<2.4\;.\label{deltaest}\end{equation}
In the case when $\mu_0=0$, we have $\sigma=0$ and hence using
(\ref{constexen}) and~(\ref{consten}), we obtain
\begin{equation}r_0=\frac{\delta}{\varepsilon\varepsilon^\ast}\varepsilon=
\frac{\sqrt{2}\delta_0}{ m_0}\frac{4\pi m_0^2}{g^2N}\sqrt{1+\sqrt{1-\left(
\frac{g^2N}{4\pi m_0^2\alpha_0}\right)^2}}\,,\label{gstrnot}\end{equation}
for the size of the ground state and
\begin{equation}r_0^\ast=\frac{\delta}{\varepsilon\varepsilon^\ast}
\varepsilon^\ast=\frac{\sqrt{2}\delta_0}{ m_0}\frac{4\pi m_0^2}{ g^2N}
\sqrt{1-\sqrt{1-\left(\frac{g^2N}{4\pi m_0^2\alpha_0}\right)^2}}\,,
\label{exstrnot}\end{equation}
for the size of the excited state.
Notice that $r_0\rightarrow\infty$ as $g/m_0\rightarrow0$ while
$r_0^\ast$ stays finite. This singularity disappears when $\mu_0\neq0$,
since in this case $g\geq g_0>0$ (see equation~(\ref{goneexpl})). For the
case of minimal masses, equation~(\ref{quantgmu}) for $\mu_0=0$, the
sizes are
\begin{equation}r_0=r_0^\ast=\frac{1}{m_0}\frac{\sqrt{2}\delta_0}{\alpha_0}\,.
\label{minmasssize}\end{equation}
For the ground state~(\ref{minmasssize}) corresponds to the minimal size,
while for the excited state - to the maximal size.

\section{Derrick's theorem and proof of local stability}\label{Derrick}

Derrick's theorem~\cite{Derrick} refers to time-independent solutions of a
class of nonlinear equations for real scalar fields and it consists of
two parts. The first part is the virial theorem and the second part is
the proof of local instability. We now recall Derrick's theorem and
discuss some of the problems connected to it. We then generalize this
theorem to a class of theories which contains the model investigated
in this paper, and derive the conditions of local stability. In
particular we prove the local stability of the ground state solution
studied in this paper.

Consider the lagrangian for a real scalar field $\theta$
\begin{equation}{\cal L}=\frac{1}{2}\left(\left(\frac{\partial\theta}{\partial
t}\right)^2-\left(\nabla\,\theta
\right)^2-f\left(\theta\right)\right)\label{Dlagr}\end{equation}
and the corresponding equation of motion
\begin{equation}\triangle\theta-\frac{\partial^2\theta}{\partial t^2}=
\frac{1}{2}f'\left(\theta\right)\,,\label{Deom}\end{equation}
where $f$ is a smooth function. A time-independent solution
$\theta\left({\bf r} \right)$ of~(\ref{Deom}) corresponds to the
extremum $\delta H=0$ of the energy functional
\begin{equation}H=\int{\rm d}^3{\bf
r}\left[\left(\nabla\,\theta\right)^2+f\left(\theta\right)\right]\equiv
I_1+I_2\,.\label{Denergy}\end{equation}
Using this fact and a particular form of the variation $\delta H$,
Derrick proved that the kinetic part $I_1$ and the potential part $I_2$
are related according to
\begin{equation}I_1+3I_2=0\,.\label{Dvir}\end{equation}
Equation~(\ref{Dvir}) constitutes the virial theorem. In the case
$f\left(\theta\right)\geq0$ this theorem precludes the existence of
time-independent solutions of~(\ref{Deom}) since in this
case both $I_2>0$ and $I_1>0$, which contradicts~(\ref{Dvir}).
If $f\left(\theta\right)\geq0$ is not valid, the energy $H$ is not
bounded from below and hence a time-independent solution
of~(\ref{Deom}) can be stable at most locally. However, using a
particular form of the variation, Derrick showed that
\begin{equation}\delta^2H=-2I_1<0\,.\label{Dinstab}\end{equation}
Local stability requires $\delta^2 H>0$ (local minimum of the total
energy) for all possible variations, but to prove the local
instability it is sufficient to show that $\delta^2 H\leq0$ for a
particular variation, so that~(\ref{Dinstab}) implies that all
time-independent solutions of~(\ref{Deom}) are locally unstable.
Equation~(\ref{Dinstab}) constitutes the second part of Derrick's
theorem. Shortly after Derrick's paper Rosen~\cite{Rosen} proved that
$\delta^2 H>0$ is the necessary and sufficient condition for dynamical
stability in the sense of Liapunov.

In the subsequent repetitions of Derrick's theorem (see for instance
ref.~\cite{Rajaraman}) the second part of Derrick's theorem was
dropped and the local stability condition $\delta^2 H>0$
replaced by the much stronger condition $f\left(\theta\right)\geq0$.
However, if one admits time-dependent solutions, in particular
stationary bound state solutions of the form
$\Psi\left({\bf r},t\right)={\rm e}^{-i Et}\psi\left({\bf r}\right)$,
then the latter condition cannot be justified in general and the
former, contrary to~(\ref{Dinstab}), can be proved now for particular
cases. Thus in particular cases at least locally stable bound state
solutions are possible. A simple illustration of this fact is provided by
\begin{equation}\left(\Box-m_0^2+gq\left(\Psi^\ast\Psi\right)^{q-1}\right)
\Psi=0\,,\quad1<q<5/3\,,\label{stexample}\end{equation}
where $\Box\equiv\triangle-\partial^2/\partial t^2$.
The local stability of stationary bound states of~(\ref{stexample})
was proved in ref.~\cite{Shatah}. Notice that the energy corresponding
to (\ref{stexample}) is
\begin{equation}H=\int{\rm d}^3{\bf r}\left[\nabla\,\Psi^\ast\cdot\nabla\,
\Psi+\frac{\partial\Psi^\ast}{\partial t}\frac{\partial\Psi}{\partial
t}+m_0^2\Psi^\ast\Psi-g\left(\Psi^\ast\Psi\right)^q\right]\,,
\label{exampleen}\end{equation}
and that the interaction potential (last term in~(\ref{exampleen})) is
strictly negative ($g$ is a positive constant) and unbounded, and
nevertheless the ground state has a finite energy and is locally
stable. The proof of local stability for the nonrelativistic analog
of~(\ref{stexample}) and other examples of stable theories with a
nonpositive interaction potential can be found in ref.~\cite{Blowup}
and references therein.

Consider now a system of $N$ complex scalar fields $\Psi_j\left({\bf
r},t\right)$, $j=1,\ldots,N$ and a real scalar field $\Phi\left({\bf
r},t\right)$ with the lagrangian
\begin{equation}{\cal L}=-\sum_{j=1}^N\left(\partial^\nu\Psi_j^\ast\partial_\nu
\Psi_j+m_0^2\Psi_j^\ast\Psi_j-g\Phi^p\left(\Psi_j^\ast\Psi_j\right)^q\right)-
\frac{1}{2}\left(\partial^\nu\Phi\partial_\nu\Phi+\mu_0^2\Phi^2\right)
\label{stablagr}\end{equation}
and the equations of motion
\begin{equation}\left(\Box-m_0^2+gq\Phi^p\left(\Psi_j^\ast\Psi_j\right)^{q-1}
\right)\,\Psi_j=0\,,\label{stabeom}\end{equation}
\begin{equation}\left(\Box-\mu_0^2\right)\,\Phi=-gp\Phi^{p-1}\sum_{i=1}^N\left(
\Psi_i^\ast\Psi_i\right)^q\,,\label{stabeomU}\end{equation}
where $m_0$, $g$, $p$ and $q$ are real positive constants and the fields
$\Psi_j$ are normalized according to
\begin{equation}\int{\rm d}^3{\bf r}\left|\frac{\partial\Psi^\ast_j}{\partial
t}\Psi_j-\Psi^\ast_j\frac{\partial\Psi_j}{\partial t}\right|=1\,,\quad j=1,
\ldots,N\,.\label{stabnorm}\end{equation}
Notice that the left-hand side of~(\ref{stabnorm}) is a constant of motion.
There are four important special cases to notice: $p=q=1$ gives
the model investigated in this paper, $p=0$ yields the case of
(\ref{stexample}), and $q=0$, $p=3$ yields the standard $\Phi^3$ field theory,
while $q=0$, $p=4$ yields the standard $\Phi^4$ field theory.
The local stability condition for the second case has been derived in
ref.~\cite{Shatah} and the existence of stable bound state solutions in the
latter two cases has been ruled out already by Derrick's theorem. We
now generalize Derrick's theorem to the class of theories characterized
by~(\ref{stablagr}-\ref{stabnorm}) for the case of time-dependent but
stationary bound state solutions of the form
$\Psi_j\left({\bf r},t\right)=\psi_j\left({\bf r}\right)\,{\rm e}^{-iE_jt}$,
$\Phi\left({\bf r},t\right)=\,\phi\left({\bf r}\right)$ and for the
case of arbitrary $p$ and $q$. Since this class is less general than
in the case of Derrick's theorem, we will benefit by being able to
derive two additional virial relations.

For stationary fields the equations of motion~(\ref{stabeom}),~(\ref{stabeomU})
and the normalization condition~(\ref{stabnorm}) become respectively
\begin{equation}\left(\triangle-\gamma_j^2+gq\phi^p\left(\psi_j^\ast\psi_j\right)^{q-1}\right)\psi_j
=0\,,\label{stabHartreeeq}\end{equation}
\begin{equation}\left(\triangle-\mu_0^2\right)\phi=-gp\phi^{p-1}\sum_{i=1}^N\left(\psi_i^\ast
\psi_i\right)^q\,,\label{stabHartreepoteq}\end{equation}
and
\begin{equation}<\psi_j\vert\psi_j>\equiv2\vert E_j\vert\int{\rm d}^3{\bf r}
\left|\psi_j\left({\bf r}\right)\right|^2=1\,,\quad
j=1,\ldots,N\,,\label{stabnormcond}\end{equation}
where $\gamma^2_j=m_0^2-E_j^2>0$. The corresponding total energy is
\begin{eqnarray}&&H=\int{\rm d}^3{\bf r}\left(\sum_{j=1}^N\left[\nabla
\psi_j^\ast\cdot\nabla\psi_j+\left(E_j^2+m_0^2\right)\psi_j^\ast\psi_j-g
\phi^p\left(\psi_j^\ast\psi_j\right)^q\right]+\frac{1}{2}\left[\left(\nabla
\phi\right)^2+\mu_0^2\phi^2\right]\right)\nonumber\\
&&\hspace{0.5truecm}\equiv H_1+H_2-H_I+H_3+H_4\,.\label{staben}\end{eqnarray}
Notice that~(\ref{stabnormcond}) implies
\begin{equation}H_2=\sum_{j=1}^N\left[\vert
E_j\vert+\frac{m_0}{2}\left(\frac{m_0}{\vert E_j\vert}-
\frac{\vert E_j\vert}{ m_0}\right)\right]\,.\label{stabentwo}\end{equation}
\par
In order to make the variation of the energy $\delta H$ we have to choose
a proper energy functional. Equations~(\ref{stabnormcond}) tell us that we
are dealing with a constrained system. According to the standard rules of
quantum mechanics there are two ways to perform variation of the
energy $\delta H$ for a constrained system. The first~\cite{Messiah}
is to use an energy functional which does not depend on the norm
of the variational functions corresponding to $\psi_j$. The
second~\cite{Bethe} is to introduce the normalization condition by real
Lagrange multipliers. Both ways are equivalent, but for our purposes
the first is more convenient. The unique functional of the
variational fields $\psi_{\lambda j}$ and $\phi_\lambda$, which
fulfills the above requirement and which reduces to~(\ref{staben}) in the case
$\psi_{\lambda j}=\psi_j$, $\phi_\lambda=\phi$ is
\begin{eqnarray}H\left(\lambda\right)=&\int{\rm d}^3{\bf
r}\sum_{j=1}^N\left[\frac{\nabla\,
\psi_{\lambda j}^\ast\cdot\nabla\,\psi_{\lambda j}+\left(E_j^2+m_0^2\right)
\psi_{\lambda j}^\ast\psi_{\lambda j}}{<\psi_{\lambda j}\vert
\psi_{\lambda j}>}-g\phi_\lambda^p\frac{\left(\psi_{\lambda j}^\ast
\psi_{\lambda j}\right)^q}{<\psi_{\lambda j}\vert\psi_{\lambda j}>^q}
\right]\\
&+\frac{1}{2}\int{\rm d}^3{\bf
r}\left[\left(\nabla\,\phi_\lambda\right)^2+\mu_0^2
\phi_\lambda^2\right]\,.\label{energyfunc}\end{eqnarray}
Now to perform the variation $\delta H$ we have to choose a set of
variational fields $\psi_{\lambda j}$ and $\phi_\lambda$, which is
large enough to yield all the solutions of $\delta H=0$ or,
equivalently, the equations of motion~(\ref{stabHartreeeq})
and~(\ref{stabHartreepoteq}). A suitable set of variational fields is
\begin{equation}\psi_{\lambda j}\left({\bf
r}\right)=\lambda^t\psi\left(\lambda{\bf r}\right)\,,\quad
\psi_{\lambda j}^\ast\left({\bf
r}\right)=\lambda^{t^\ast}\psi^\ast\left(\lambda{\bf r}
\right)\,,\quad\phi_\lambda\left({\bf r}\right)=\lambda^s\phi\left(\lambda{\bf
r}\right)\,,
\label{varfields}\end{equation}
where $t$ is an arbitrary complex and $\lambda$ and $s$ are arbitrary
real numbers. The variational fields $\psi_{\lambda j}$ are not
normalized except for $\lambda=1$ in which case the normalization is
defined in~(\ref{stabnormcond}). For $\lambda=1$ we have
\begin{equation}\psi_{1j}\left({\bf r}\right)=\psi_j\left({\bf
r}\right)\,,\quad\psi_{1j}^\ast\left({\bf r}\right)=
\psi_j^\ast\left({\bf r}\right)\,,\quad\phi_1\left({\bf
r}\right)=\phi\left({\bf r}\right)\,,\quad
H\left(1\right)=H\,.\label{varfieldsone}\end{equation}
The variations $\delta\psi_j$, $\delta\psi_j^\ast$, $\delta\phi$,
$\delta H$ and $\delta^2H$ are defined by
\begin{equation}\delta\psi\equiv\left(\frac{{\rm d}\psi_{\lambda j}}{{\rm
d}\lambda}
\right)_{\lambda=1}=t\,\psi\left({\bf r}\right)+{\bf
r}\cdot\nabla\psi\left({\bf r}\right)\,,\quad
\delta\psi^\ast\equiv\left(\frac{{\rm d}\psi_{\lambda j}^\ast}{{\rm d}\lambda}
\right)_{\lambda=1}=t^\ast\,\psi^\ast\left({\bf r}\right)+{\bf
r}\cdot\nabla\psi^\ast\left(
{\bf r}\right)\label{varfieldstwo}\end{equation}
and
\begin{equation}\delta\,\phi\equiv\left(\frac{{\rm d}\phi_\lambda}{{\rm
d}\lambda}
\right)_{\lambda=1}=s\,\phi\left({\bf r}\right)+{\bf
r}\cdot\nabla\phi\left({\bf r}\right)\,,\quad
\delta H=\left(\frac{{\rm d}H\left(\lambda\right)}{{\rm
d}\lambda}\right)_{\lambda=1}\,,\quad
\delta^2H=\left(\frac{{\rm d}^2H\left(\lambda\right)}{{\rm
d}\lambda^2}\right)_{\lambda=1}\,.
\label{varfieldsthree}\end{equation}
Notice that at each ${\bf r}$ the variations $\delta\psi_j$,
$\delta\psi_j^\ast$ and $\delta\phi$ are arbitrary and independent of
each other. Therefore, applying the standard variational procedure of
quantum mechanics, we conclude that $\delta H=0$ is equivalent to
(\ref{stabHartreeeq}-\ref{stabHartreepoteq}).
\par
Since the integration in~(\ref{energyfunc}) is over the entire space, we
can eliminate the dependence of the fields on $\lambda{\bf r}$ by making a
suitable rescaling of the integration variable and obtain
\begin{equation}H\left(\lambda\right)=\lambda^2H_1+H_2-\lambda^{sp+3\left(q-1\right)}H_I+\lambda^{2s-1}
H_3+\lambda^{2s-3}H_4\,,\label{enlambda}\end{equation}
where $H_1,\ldots,H_4$ and $H_I$, which are independent of $\lambda$,
are defined in~(\ref{staben}). From~(\ref{enlambda}) and $\delta H=0$ it
follows
that
\begin{equation}2H_1-\left[sp+3\left(q-1\right)\right]H_I+\left(2s-1\right)
H_3+\left(2s-3\right) H_4=0\,.
\label{Derone}\end{equation}
Since~(\ref{Derone}) must be valid for all $s$, it implies two separate
virial relations
\begin{equation}2H_3+2H_4-pH_I=0\label{Dertwo}\end{equation}
and
\begin{equation}2H_1-H_3-3H_4-3\left(q-1\right)
H_I=0\,.\label{Derthree}\end{equation}
Moreover, multiplying~(\ref{stabHartreeeq}) by $\psi_j^\ast$, integrating
over the entire space and summing over all $j=1,\ldots,N$, we obtain a
third independent virial relation
\begin{equation}H_1+H_2-qH_I=\sum_{j=1}^N\vert
E_j\vert\,.\label{Derfour}\end{equation}
Combining these relations one can obtain other useful relations.
For instance, eliminating $H_1$ and $H_I$ and using~(\ref{stabentwo}) we
obtain
\begin{equation}\left[2\left(3-q\right)-p\right]H_3+\left[2\left(3-q\right)-
3p\right]H_4=m_0\sum_{j=1}^N\left(\frac{m_0}{\vert E_j\vert}-\frac{\vert E_j
\vert}{ m_0}\right)\,.\label{stabenthree}\end{equation}
Since $H_3$, $H_4$ and the right-hand side of~(\ref{stabenthree}) are
positive numbers, we obtain a necessary condition for the existence of
bound states
\begin{equation}2\left(3-q\right)-p>0\,.\label{Dexcond}\end{equation}
Another useful relation is obtained if we eliminate $H_3$ from
(\ref{stabenthree}) by means of~(\ref{Dertwo})
\begin{equation}\frac{p}{2}\left[2\left(3-q\right)-p\right]H_I=m_0\sum_{j=1}^N
\left(\frac{m_0}{\vert E_j\vert}-\frac{\vert E_j\vert}{m_0}\right)+2pH_4\,.
\label{anotherrel}\end{equation}
This is the equation~(\ref{pmvirtheor}) which we used above to determine the
mass.
\par
Now we compute the second variation of the energy $\delta^2H$ and
derive the local stability conditions which ensure that $\delta^2H>0$ for
all values of $s$ and $t$, which are the parameters spanning the set of
variations. Equation~(\ref{enlambda}) implies
\begin{eqnarray}&&\hspace{-1truecm}\delta^2H=2H_1-\left[sp+3\left(q-1\right)
\right]\left[sp+3\left(q-1\right)-1\right]H_I+\left(2s-1\right)\left(2s-2
\right)H_3\nonumber\\
&&\hspace{9truecm}+\left(2s-3\right)\left(2s-4\right) H_4\,.
\label{Derfive}\end{eqnarray}
Eliminating $H_1$, $H_3$ and $H_I$ by means of the virial relations
(\ref{Dertwo}),~(\ref{Derthree}) and~(\ref{stabenthree}), we obtain
\begin{eqnarray}&&\hspace{-1.7truecm}\delta^2H=\frac{2p\left(2-p
\right)s^2-4p\left(3q-2\right)s+3p-6\left(q-1\right)\left(3q-5\right)}
{2\left(3-q\right)-p}m_0\sum_{j=1}^N\left(\frac{m_0}{\vert E_j\vert}-
\frac{\vert E_j\vert}{m_0}\right)\nonumber\\
&&\hspace{-1.7truecm}+\frac{2p\left(2-p\right)s^2-\left\{12p\left(
q-1\right)+8\left(3-q\right)\right\}s-3p+12\left(3-q\right)-6\left(q-1\right)
\left(3q-5\right)}{2\left(3-q\right)-p}2H_4\,.\label{Dersix}\end{eqnarray}
Consider the case when $\mu_0=0$. From~(\ref{staben}) we have $H_4=0$, and
hence~(\ref{Dersix}) implies that $\delta^2 H>0$ is equivalent to
\begin{equation}2p\left(2-p\right) s^2-4p\left(3q-2\right)s+3p-6\left(q-1
\right)\left(3q-5\right)>0\,,\quad\mu_0=0\,.
\label{Derseven}\end{equation}
The solution of this inequality is
\begin{equation}\frac{4-p}{3}-\frac{\vert2-p\vert}{6}<q<\frac{4-p}{3}+
\frac{\vert2-p\vert}{6}\,,\quad\mu_0=0\,.\label{Dereight}\end{equation}
Notice that for $p=0$~(\ref{Dereight}) coincides with the local stability
condition proved in ref.~\cite{Shatah}, which we quoted in equation
(\ref{stexample}). Also notice that the only positive integers $p$ and
$q$ which can satisfy~(\ref{Dereight}) are $p=1$ and $q=1$, which is
the case investigated in this paper. To have a sensible field theory
$p$ and $q$ must be positive integers. Therefore the result - that
stability alone restricts the choice among the class of theories
defined by~(\ref{stablagr}) to just one case: $p=q=1$ - must be
considered as satisfactory.

For each $p$ and $q$ obeying~(\ref{Dereight}), the local stability for
a sufficiently small $\mu_0\neq0$ follows by continuity of $H_4$. The
local stability condition $\delta^2 H>0$ implies an upper bound on $\mu_0$.
For instance, for $p=1$ and $q=1$ we substitute equation~(\ref{phibeta})
in~(\ref{Dersix}), which in the present notation reads
\begin{equation}H_4=\beta m_0N\left(\frac{m_0}{\vert E\vert}-\frac{\vert E\vert
}{ m_0}\right)\,,\label{Dernine}\end{equation}
and obtain the local stability condition for the ground state
\begin{equation}s^2-2\frac{1+8\beta}{1+2\beta}s+\frac{3}{2}\frac{1+14\beta}
{1+2\beta}>0\,,\label{Derten}\end{equation}
which must be satisfied for all real $s$. Solving for $\beta$, we obtain
\begin{equation}\beta<\frac{2}{11}\left(1+\frac{3\sqrt{3}}{4}\right)\,,
\label{Dereleven}\end{equation}
and then using~(\ref{betaappr}) and~(\ref{pmethree})
\begin{equation}\sigma<1.0\,,\quad\mu_0<\sqrt{2}m_0\,.
\label{Dertwelve}\end{equation}
Thus a locally stable ground state for the theory investigated in this
paper ($p=q=1$) exists only if $\sigma<1$.

\section*{Acknowledgments}

Most of this work was done at the Weizmann Institute of Science, Rehovot.
The author wishes to thank Y. Eisenberg, S. A. Gurvitz and D. Kazhdan
for valuable discussions and suggestions.

\appendix\section*{Appendix}

In this appendix we develop a simple algorithm for the numerical
determination of the functions needed for the evaluation of the functions
$R_\sigma$ and ${\cal V}_\sigma$. As a first step we split the recurrence
relation~(\ref{pmrrelphitwo}) into three separate pieces to avoid multiple
integration
\begin{equation}\varphi_{nm}\left(y,\sigma\right)=\int_0^y{\rm d}t\;\phi_{nm}
\left(t,\sigma\right)\,,\label{pmrrphi}\end{equation}
\begin{eqnarray}\lefteqn{\phi_{nm}\left(y,\sigma\right)=\frac{1}{\left(y+
n\sigma+m+\frac{1}{2}\right)^2-\frac{1}{4\,}}\Biggl(\varphi_{n-1,m}\left(y,
\sigma\right)}\hspace{6truecm}\nonumber\\
&&\left.+\sum_{k=0}^n\sum_{l=1}^m\int_0^y{\rm d}t\varphi_{n-k,m-l}\left(y-t,
\sigma\right)f_{kl}\left(t,\sigma\right)\right)\label{pmrrphione}\end{eqnarray}
and
\begin{eqnarray}\lefteqn{f_{nm}\left(y,\sigma\right)=\frac{1}{\left(y+n\sigma+m
\right)^2-\sigma^2}\Biggl(\varphi_{n,m-1}\left(y,\sigma\right)}
\hspace{5.3truecm}\nonumber\\
&&\left.+\sum_{k=0}^n\sum_{l=1}^m\int_0^y{\rm d}t\varphi_{n-k,m-l}\left(y-t,
\sigma\right)\phi_{k,l-1}\left(t,\sigma\right)\right).
\label{pmrrphitwo}\end{eqnarray}
It is easy to see that these equations are explicit recurrence relations,
rather than integral equations. In order to convert these recurrence
relations to a form digestible by computers, we divide the interval
$\left[0,y\right]$ into ${\cal N}$ pieces each of length $x$ and use
the trapezoidal rule to evaluate the integrals. As a result of the
discretization we obtain a new set of functions
$\overline{\varphi}_{nm}\left(x_j,\sigma\right)$,
$\overline{\phi}_{nm}\left(x_j,\sigma\right)$ and
$\overline{ f}_{nm}\left(x_j,\sigma\right)$ defined on the grid of points
$x_0\equiv0,x_1,\ldots,x_{\cal N}\equiv y$, which converge to the
true functions $\varphi_{nm}\left(x_j,\sigma\right)$,
$\phi_{nm}\left(x_j,\sigma\right)$ and $ f_{nm}\left(x_j,\sigma\right)$ as
$x\rightarrow0$. The corresponding recurrence relations are
\begin{equation}\overline{\varphi}_{nm}\left(x_j,\sigma\right)=\,\overline
{\varphi}_{nm}\left(x_{j-1},\sigma\right)+\frac{x}{2}\left(\,\overline
{\phi}_{nm}\left(x_{j-1},\sigma\right)+\,\overline{\phi}_{nm}\left(x_j,\sigma
\right)\right)\,,\label{pmrrphitra}\end{equation}
\begin{eqnarray}&&\hspace{-1.2truecm}\overline{\phi}_{nm}\left(x_j,\sigma
\right)=\frac{1}{\left(x_j+n\sigma+m+\frac{1}{2}\right)^2-\frac{1}{4\,}}
\Biggl({\overline{\varphi}_{n-1,m}}\left(x_j,\sigma\right)+\frac{x}{2}\,
\overline{f}_{nm}\left(x_j,\sigma\right)+\frac{x}{2}\,\frac{\overline{
\varphi}_{n,m-1}\left(x_j,\sigma\right)}{1-\sigma^2}\nonumber\\
&&\hspace{5.8truecm}\left.+\,x\sum_{k=0}^n\sum_{l=1}^m\sum_{i=1}^{j-1}
\overline{\varphi}_{n-k,m-l}\left(x_j-x_i,\sigma\right)\,\overline{f}_{kl}
\left(x_i,\sigma\right)\right)\label{pmrrphionetra}\end{eqnarray}
and
\begin{eqnarray}&&\hspace{-1.1truecm}\overline{f}_{nm}\left(x_j,\sigma\right)
=\frac{1}{\left(x_j+n\sigma+m\right)^2-\sigma^2}\left({\overline{\varphi}_{n,
m-1}}\left(x_j,\sigma\right)+\frac{x}{2}\,\overline{\phi}_{n,m-1}\left(x_j,
\sigma\right)+\frac{x}{2}\,\frac{\overline{\varphi}_{n-1,m-1}\left(x_j,\sigma
\right)}{\sigma\left(\sigma+1\right)}\right.\nonumber\\
&&\hspace{5.3truecm}\left.+\,x\sum_{k=0}^n\sum_{l=1}^m\sum_{i=1}^{j-1}
\overline{\varphi}_{n-k,m-l}\left(x_j-x_i,\sigma\right)\,\overline{\phi}_{k,
l-1}\left(x_i,\sigma\right)\right),\label{pmrrphitwotrap}\end{eqnarray}
where $x_j=jx$ and $j=1,\ldots,{\cal N}$, $n=0,1,\ldots,$
$m=0,1,\ldots.$ The initial data for the recurrence process are
\begin{equation}\overline{\varphi}_{-1,m}\left(x_j,\sigma\right)=0\,,\quad
\overline{\varphi}_{n,-1}\left(x_j,\sigma\right)=0\,,\quad\overline
{\varphi}_{nm}\left(0,\sigma\right)=0\,,\label{trapinitcond}\end{equation}
except for $n=m=0$ in which case
\begin{equation}\overline{\varphi}_{00}\left(0,\sigma\right)=1\,,
\label{trapinitcondone}\end{equation}
and
\begin{equation}\overline{\phi}_{n,-1}\left(x_j,\sigma\right)=0\,,\quad
\overline{\phi}_{nm}\left(0,\sigma\right)=0\,,\label{trapinitcondtwo}
\end{equation}
except for $n=1$, $m=0$ in which case
\begin{equation}\overline{\phi}_{10}\left(0,\sigma\right)=\frac{1}{\sigma
\left(\sigma+1\right)}\,.\label{trapinitcondthree}\end{equation}
The recurrence process starts with the evaluation of the cycle
(\ref{pmrrphitwotrap}) $\rightarrow$ ~(\ref{pmrrphionetra}) $\rightarrow$
(\ref{pmrrphitra}) for $n=m=0$ and $j=1$. Then $n$, $m$ and $j$ are iterated
until certain maximal values, $j_{\rm max}$, $n_{\rm max}$ and
$m_{\rm max}$ say, are reached, which are determined by the
step-function in~(\ref{pmAnsR}). The result of the recurrence process is
$\overline{\varphi}_{nm}\left(x_1,\sigma\right),\ldots,\overline{\varphi}_{nm}
\left(x_{j_{\rm max}},\sigma\right)$, $n=0,\ldots,n_{\rm max}$,
$m=0,\ldots,m_{\rm max}$. In order to obtain an estimate of the
difference between $\overline{\varphi}_{nm}\left(x_j,\sigma\right)$
and $\varphi_{nm}\left(x_j,\sigma\right)$, the grid is refined by replacing
${\cal N}$ by $2{\cal N}$ and $x$ by $x/2$, and the recurrence process
is repeated. The resulting $\overline{\varphi}_{nm}\left(x_2,\sigma\right),
\overline{\varphi}_{nm}\left(x_4,\sigma\right),\ldots$ are compared with the
previously calculated $\overline{\varphi}_{nm}\left(x_1,\sigma\right),
\overline{\varphi}_{nm}\left(x_2,\sigma\right),\ldots$. The process terminates
when a certain specified accuracy is reached and the $\overline{\varphi}_{nm}
\left(x_1,\sigma\right),\overline{\varphi}_{nm}\left(x_2,\sigma\right),\ldots$
can be considered to be identical to $\varphi_{nm}\left(x_1,\sigma\right),
\varphi_{nm}\left(x_2,\sigma\right),\ldots.$ Notice that the numerical
evaluation of the functions $\varphi_{nm}$ does not mean that we are
solving the problem numerically. Rather, it means that the solution
(\ref{pmAnsR}) is given in terms of nonstandard functions and that we have to
teach our computer to obtain the values of these well-defined analytical
functions.

\listoffigures
\begin{figure}[b]
\caption{a) The mass to bare mass ratios $M_N/Nm_0$ (solid line),
$M_{0NN}/Nm_0$ (dashed line) and $M_{N0N}/Nm_0$ (dotted-dashed line) as
functions of $g^2N/4\pi m_0^2$ for various fixed values of
$\mu_0/m_0$. Dotted vertical lines correspond to the maximal values of
$g^2N/4\pi m_0^2$, the dotted horizontal line corresponds to the
absolute minimum of the mass $M/Nm_0=2\protect\sqrt{2}/3$, which
occurs at $\mu=0$.}\label{MassN}
\end{figure}
\addtocounter{figure}{-1}
\begin{figure}[b]
\caption{b) The mass to bare mass ratios $M/Nm_0$ (solid line),
$M_{0NN}/Nm_0$ (dashed line) and $M_{N0N}/Nm_0$ (dotted-dashed line) as
functions of $\sigma$ for various fixed values of $\mu_0/m_0$. Dotted
vertical lines correspond to the maximal values of $g^2N/4\pi m_0^2$
($\sigma=\sigma_1$), the dotted horizontal line corresponds to the
absolute minimum of the mass $M/Nm_0=2\protect\sqrt{2}/3$, which occurs at
$\mu=0$.}
\end{figure}
\begin{figure}[b]
\caption
{a) The wave function $\psi\left(r\right)$ plotted as a function of
the dimensionless variable $\gamma r$ for various values of the parameter
$\sigma=\mu_0/2\gamma$, where $\gamma=\left(m_0^2-E^2\right)^{1/2}$.
The function  is scaled by an appropriate factor in order to become
a dimensionless wave function of only two variables $\gamma r$ and $\sigma$.
The latter depends only on the fundamental parameters of the model,
such as the coupling constant $g$.}\label{wfplot}
\end{figure}
\addtocounter{figure}{-1}
\begin{figure}[b]
\caption
{b) The same as in figure~\protect\ref{wfplot}a for the potential
function $-\phi\left(r\right)$.}
\end{figure}
\begin{figure}[b]
\caption{a) The function $R_\sigma\left(\eta,\zeta,s\right)$
as a function of $s$ for $\sigma=0.1$ and $\eta=2.448(5)$, $\zeta=582.(3)$.
The constants $\eta$ and $\zeta$ were determined from the boundary conditions
(\protect\ref{bstcondR}) and (\protect\ref{pmeta}).}\label{RVplot}
\end{figure}
\addtocounter{figure}{-1}
\begin{figure}[b]
\caption{b) The same as in figure \protect\ref{RVplot}a for the
function ${\cal V}_\sigma\left(\eta,\zeta,s\right)$.}
\end{figure}
\begin{figure}[b]
\caption{The constituent energies $\varepsilon/m_0$ and
$\varepsilon^\ast/m_0$ as functions of the coupling constant
$g^2N/4\pi m_0^2$ at various fixed values of $\mu_0/m_0$. The dotted line
indicates the value $1/\protect\sqrt{2}$ which is the boundary value
deviding the two functions.}\label{eofgn}
\end{figure}
\begin{figure}[b]
\caption{$g^2N/4\pi m_0^2$ as a function of $\sigma$
for various values of $\mu_0/m_0$ (solid lines). The vertical dotted
lines and the vertical dashed line give the values of $\sigma_1$ and
$\sigma_0$ for the maximum and the minimum of $g$ respectively, according
to the approximate formulae given in the text.}\label{gNofsig}
\end{figure}
\end{document}